\PassOptionsToPackage{unicode}{hyperref}
\PassOptionsToPackage{naturalnames}{hyperref}
\documentclass[fleqn,usenatbib]{mnras}

\usepackage{newtxtext,newtxmath}
\usepackage{txfonts}

\usepackage[T1]{fontenc}

\DeclareRobustCommand{\VAN}[3]{#2}
\let\VANthebibliography\thebibliography
\def\thebibliography{\DeclareRobustCommand{\VAN}[3]{##3}\VANthebibliography}

\usepackage{graphicx}	
\usepackage{amsmath}
\usepackage{amsfonts}

\usepackage{xcolor}
\usepackage{amsmath}

\definecolor{vg}{rgb}{0.55, 0.0, 0.55}

\title[DM unIDs with ML]{A search for dark matter among Fermi-LAT unidentified sources with systematic features in Machine Learning}

\author[V. Gammaldi et al.]{
V. Gammaldi,$^{1,2}$\thanks{E-mail: viviana.gammaldi@uam.es}
B. Zaldívar,$^{3}$
M. A. Sánchez-Conde$^{1,2}$
and J. Coronado-Blázquez$^{1,2,4}$
\\
$^{1}$Departamento de F\'isica Te\'orica, Universidad Aut\'onoma de Madrid, Cantoblanco, 28049, Madrid, Spain\\
$^{2}$ Instituto de F\'isica Te\'orica, IFT-UAM/CSIC, Cantoblanco, 28049, Madrid, Spain\\
$^{3}$ Institute of Corpuscular Physics (IFIC), University of Valencia and CSIC, Calle Catedr\'atico Jos\'e Beltr\'an 2, 46980 Paterna, Spain\\
$^{4}$Telef\'onica Tech IoT $\&$ Big Data, Ronda de la Comunicaci\'on s/n, Madrid, Spain}

\date{Accepted XXX. Received YYY; in original form ZZZ}

\pubyear{2015}

\begin{document}
\label{firstpage}
\pagerange{\pageref{firstpage}--\pageref{lastpage}}
\maketitle

\begin{abstract}
Around one third of the point-like sources in the Fermi-LAT catalogs remain as unidentified sources (unIDs) today. Indeed, these unIDs lack a clear, univocal association with a known astrophysical source. If dark matter (DM) is composed of weakly interacting massive particles (WIMPs), there is the exciting possibility that some of these unIDs may actually be DM sources, emitting gamma rays from WIMPs annihilation. We propose a new approach to solve the standard, Machine Learning (ML) binary classification problem of disentangling prospective DM sources (simulated data) from astrophysical sources (observed data) among the unIDs of the 4FGL Fermi-LAT catalogue. We artificially build two {\it systematic} features for the DM data which are originally inherent to observed data: the detection significance and the uncertainty on the spectral curvature. We do it by sampling from the observed population of unIDs, assuming that the DM distributions would, if any, follow the latter. We consider different ML models: Logistic Regression, Neural Network (NN), Naive Bayes and Gaussian Process, out of which the best, in terms of classification accuracy, is the NN, achieving around $93.3\% \pm 0.7\%$ performance. Other ML evaluation parameters, such as the True Negative and True Positive rates, are discussed in our work. Applying the NN to the unIDs sample, we find that the degeneracy between some astrophysical and DM sources can be partially solved within this methodology. Nonetheless, we conclude that there are no DM source candidates among the pool of 4FGL Fermi-LAT unIDs.

\end{abstract}




\section{Introduction}

Astrophysical and cosmological evidence suggests that non-baryonic cold DM constitutes $84\%$ of the matter density of the Universe \citep{Ade:2015xua, Aghanim:2018eyx}. Although the nature of DM is still unknown,  Weakly Interacting Massive Particles (WIMPs) are popular and well-motivated DM candidates, among others. In particular, WIMPs are one of the most popular types of DM candidates in the context of DM searches. The WIMP paradigm invokes the same thermal decoupling, which is enormously successful at making detailed predictions for many observables in the early Universe, including the abundances of light elements and the CMB \citep{1991Natur.352..769P}. Indeed, it is somewhat natural to invoke a similar paradigm to infer the abundance of DM as a thermal relic from the early Universe. The reason is that, in order to fill all the DM content that we observe in the universe, WIMPs should have sizeable interactions with the Standar Model (SM) sector, thus ensuring a rich phenomenology while still having a relevant part of their parameter space allowed from all the available experimental data. In the so-called indirect detection searches, and in particular those relying on gamma-ray measurements in the Fermi-LAT's energy range, WIMPs are the natural candidates to consider given their expected masses. In fact, WIMPs are predicted to annihilate or decay into SM particles, whose decay and hadronization processes would produce secondary particles, such as cosmic rays, neutrinos and gamma rays \citep{BuckleyHooper10, Zechlin+12, ZechlinHorns12, Belikov2012, BerlinHooper14, Bertoni+15, Bertoni+16, Schoonenberg+16, Calore+17, HooperWitte17}.  The flux of secondary particles may be observed in ground-based or satellite observatories, laying the groundwork for the indirect searches for DM. Only an agreement of several hints in the observed flux of different messengers - i.e. the multi-messenger detection- would result in a competitive claim of the indirect detection of DM \citep{Bergstrom2013, MiniReview}. This includes the issue of disentangling the DM signal from the emission of well-known astrophysical sources or diffuse astrophysical background. Indeed, DM-dominated systems - e.g. dwarf galaxies, galaxy clusters as well as the Galactic center - are benchmark targets for indirect searches for DM ( see e.g. \citet{Charles:2016pgz, Conrad:2017pms, Gammaldi:2021zdm} and refs therein). 
Among others, gamma rays are considered to be the golden messenger: they are (very-) high-energy neutral particles traveling practically undeflected along straight paths in the local Universe. \\
The Large Area Telescope (LAT) on board the NASA \textit{Fermi} satellite (\textit{Fermi}-LAT) \citep{Atwood2013} has collected more than 13 years of gamma-ray data of the full sky. Still in operation, \textit{Fermi}-LAT is a pair conversion telescope capable to observe gamma-ray photons from energies $\sim$20 MeV up to $\sim$ TeV. Several point-source catalogs have been released and contain thousands of gamma-ray objects, many of them previously unknown \citep{3FGL_paper,2FHL_paper,3FHL_paper}.  Interestingly, around one third of the point-like gamma-ray sources in the 4FGL \textit{Fermi}-LAT catalog \citep{Fermi-LAT:2019yla} as well as in other gamma-ray ground based telescopes (see e.g. \citep{MAGIC:2019vba}) remain as unidentified (unIDs) today. These unIDs lack a clear, univocal association with a known astrophysical source.\\
In the last few years, Machine Learning (ML) techniques have been applied to many different fields of astrophysics and cosmology; e.g. applied to the so-called Galactic Centre Excess (\cite{Caron:2017udl}), to the search for dark matter in dwarf galaxies (\cite{Calore:2018sdx, Alvarez:2020cmw}), as well as classification algorithms that have been applied to the \textit{Fermi}-LAT catalogues (see e.g. \citet{Mirabal2016, Bartels:2018qgr,  Kovacevic:2019wpy, Hui:2020cmv, 2020arXiv200110523V, Germani:2021zzg, Bhat:2021wtb} and references therein). The latter works have been focused on classifying unIDs as different types of known astrophysical sources (e.g Active Galactic Nuclei, pulsars, blazars). Nonetheless, if DM is made of WIMPs, there is also the exciting possibility that some of these unIDs may actually be DM sources, emitting gamma rays by WIMPs annihilation \citep{Bertone:2005xv}. In fact, the nature of DM still represents an open question in physics and cosmology, and many efforts have been devoted to understand its nature via the application of novel ML techniques in several related fields\footnote{See e.g. darkmachines.org} (e.g. \citet{Bertone:2017adx, Morice-Atkinson:2017jik, 2018MNRAS.478.3410A, Ullmo:2020lok, Feickert:2021ajf, Spencer:2021ypz, 2022MNRAS.513.1972H, BAZAROV2022100667}).\\
\\ Around one third of the point-like sources in the Fermi-LAT catalogs remain as unidentified sources (unIDs) today. Indeed, these unIDs lack a clear, univocal association with a known astrophysical source. If dark matter (DM) is composed of weakly interacting massive particles (WIMPs), there is the exciting possibility that some of these unIDs may actually be DM sources, emitting gamma rays from WIMPs annihilation. We propose a new approach to solve the standard, Machine Learning (ML) binary classification problem of disentangling prospective DM sources (simulated data) from astrophysical sources (observed data) among the unIDs of the 4FGL Fermi-LAT catalogue. Concretely, we artificially build two systematic features for the DM data which are originally inherent to observed data: the detection significance and the uncertainty on the spectral curvature. We do it by sampling from the observed population of unIDs, assuming that the DM distributions would, if any, follow the latter. We consider different ML models: Logistic Regression, Neural Network (NN), Naive Bayes and Gaussian Process, out of which the best, in terms of classification accuracy, is the NN, achieving around 93.3
In this work, we propose a new approach to solve the binary classification problem of disentangling prospective DM-source candidates from astrophysical sources among the unIDs in the 4FGL \textit{Fermi}-LAT catalogue. We work on the derived parameter space defined by the energy-peak $E_{\rm peak}$ and curvature $\beta$ of the gamma-ray spectra of source in the catalogue: the so-called \textit{Fermi}-LAT $\beta$-plot \citep{Coronado-Blazquez:2019pny}.  The observational $\beta$-plot -composed of both identified and unidentified gamma-ray sources- will be here enriched by theoretically-based DM parameters and (hereafter the so-called "DM-$\beta$" plot).

Many works have pointed out the spectral confusion between pulsars and DM annihilation signals in gamma rays (e.g., \citet{fermi_dm_satellites_paper, Mirabal2013, Mirabal2016}), which is especially relevant when considering light, $\mathcal{O}\left(10 {\rm GeV}\right)$  WIMPs, and hadronic annihilation channels such as $b\bar{b}$. Indeed, such a degeneracy is pictured as an overlapping region in the $E_{\rm peak}-\beta$ plane. Nonetheless, in our work we will show that WIMP candidates cover a broader region in this parameter space. Hereafter, we refer to the parameters of such a plot as \textit{features}, by using the benchmark ML nomenclature. Because of the present degeneracy in the $E_{\rm peak}-\beta$ plane, we introduce two \textit{systematic} features for the DM sample, motivated by 
the systematic uncertainty of the \textit{Fermi}-LAT detector, which would affect the detection of any DM source. This allow us to training the classification algorithms with four features (4F) instead of 2F. Furthermore, we also discuss the possibility to adopt a three-feature setup, by including the relative uncertainty on $\beta$ ($\beta_{\rm rel}$) via both a sampled gaussian distribution of the uncertainty itself (3F-A) and in the statistical model (3F-B). \\
\\

We consider four classification algorithms, namely, Logistic Regression (LR), Neural Network (NN), Na\"ive Bayes (NB) and Gaussian Process (GP). The LR and NN algorithms are built in the \texttt{Scikit-learn} library for data analysis with ML in python \citep{scikit-learn}, while we implement our own python codes for NB and GP models, the latter using \texttt{TensorFlow v1}, the open-source library for automatic differentiation and ML applications. \citep{tensorflow_developers_2022_5949169}. 

These four classification models have been selected according to their different advantages and capabilities\footnote{See \cite{bishop}, one of the standard reference books for Machine Learning.}: LR is arguably the simplest model (it gives linear decision boundaries among the classes of points) and consequently it is highly explainable, even though it requires numerical optimization. NN on the other hand is, a priori, arbitrarily expressive while at the same time being optimized very efficiently, the reason for which it is one of the most popular ML models for problems in a wide range of domains. NB is a model giving a priori a higher expressive power than LR (it can give non-linear decision boundaries) while requiring analytical optimization. Finally, a GP classifier (\cite{rasmussen:williams:2006}) offers as well a high expressivity with the added value of being a Bayesian model, allowing to report prediction uncertainties.  

This manuscript is organized as follows: in Section \ref{sec:beta_plot} we introduce the data. In Sec. \ref{systematic} we introduce the \textit{systematic} features that will be used by our algorithms. In Sec. \ref{Methodology} we introduce the methodology, with the classification algorithms and feature setups. In Sec. \ref{sec:class_acc} we compute the ``DM-vs-astrophysics'' classification accuracy for the selected algorithms under different setups. In Sec. \ref{sec:unids_class} we provide the results of the classification of unIDs with our best classifier from the previous exercise, before concluding in Sec.\ref{conclusions}. 

\section{Experimental and theoretical data}
\label{sec:beta_plot}

The recent 4FGL \textit{Fermi}-LAT catalogue \citep{Fermi-LAT:2019yla} is a collection of sources with associated gamma-ray spectra, containing important information about their nature. Somehow surprisingly, an important fraction of objects in the \textit{Fermi}-LAT catalogs, ca. 1/3 of the total, remain as unIDs, i.e., objects lacking a clear single association to a known object identified at other wavelengths, or to a well-known spectral type emitting only in gamma rays, e.g. certain pulsars. 
Among other prospective sources of gamma rays from DM annihilation events, dark satellites or subhalos in the Milky Way, with no optical counterparts, are the preferred candidates, as they are expected to exist in high number according to standard cosmology and they would not be massive enough to retain gas/stars, this way being pristine DM annihilating sources free of gamma-ray astrophysical backgrounds. 
Many authors have already investigated DM subhalos as prospective targets for indirect DM detection \citep{ BuckleyHooper10,  Zechlin+12, ZechlinHorns12, fermi_dm_satellites_paper,  Belikov2012, BerlinHooper14, 2014GrCo...20...47B, Bertoni+15, Bertoni+16, Schoonenberg+16, Calore+17,  HooperWitte17, Coronado-Blazquez:2019pny, Coronado_Blazquez2019, 2022PhRvD.105h3006C}. 

\begin{figure}
    \centering
    \includegraphics[width=1\linewidth]{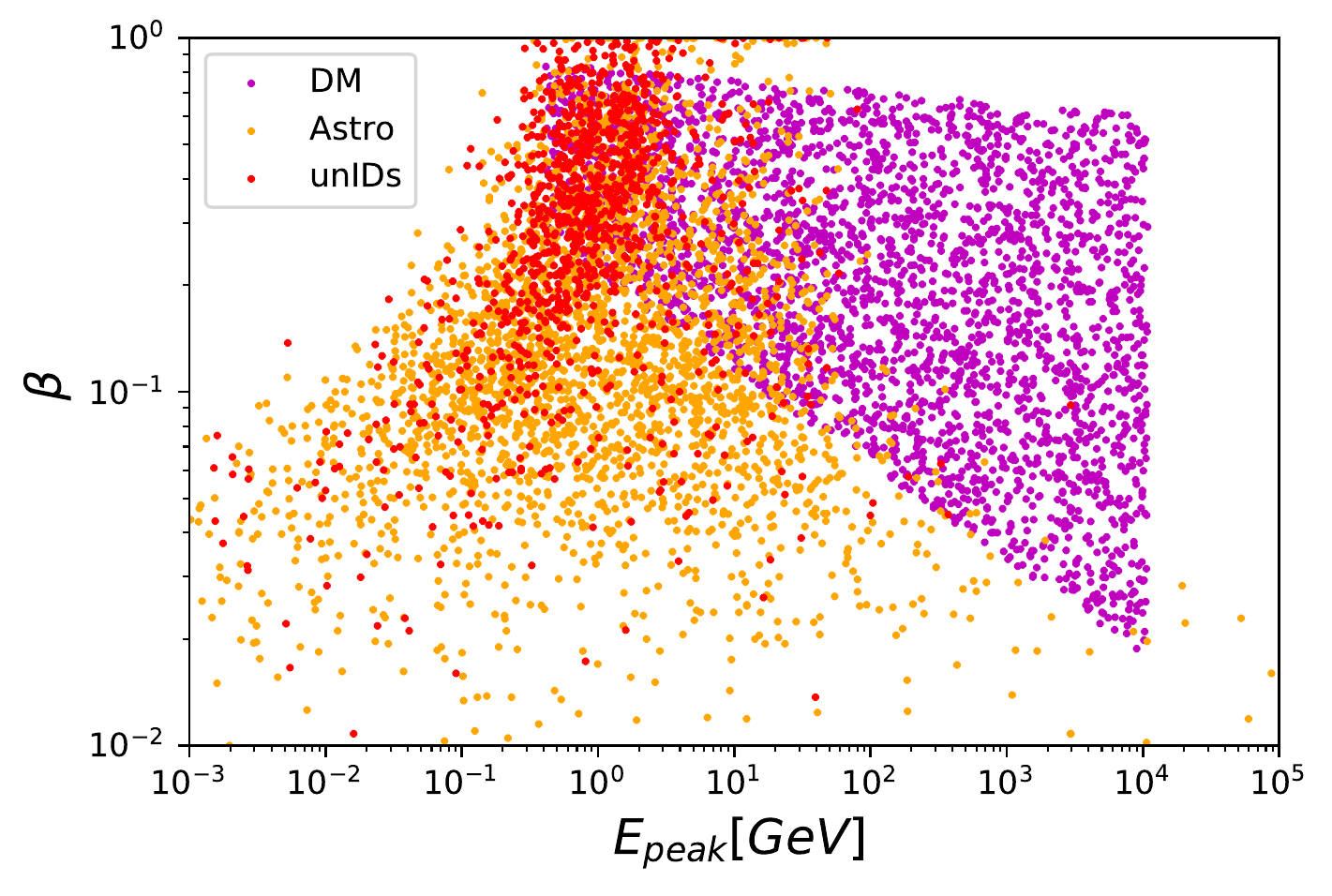}
    \caption{\protect\footnotesize{The "DM-$\beta$ plot", which includes information about the gamma-ray spectra of well-know astrophysical gamma-ray sources (orange points), unIDs sources (red points) and theoretical WIMP DM sources data set (magenta points).}}
    \label{fig:DM_beta_plot}
\end{figure}

The \textit{Fermi}-LAT 4FGL catalogue \citep{Fermi-LAT:2019yla} adopted in this work, is the result of 8 years of telescope operation. 
It covers the 50 MeV -- 1 TeV energy range, and reports the detection of over 5000 gamma-ray sources, almost doubling the previous 3FGL, and using the latest instrumental response functions (IRFs) and Pass 8 events \citep{Atwood2013}, which optimize the instrument capacities, as well as an updated Galactic diffuse emission model. In particular, we are interested in one of the parametrizations of the gamma-ray spectrum used in the 4FGL, known as the Log-Parabola (LP): 

\begin{equation}
\frac{dN}{dE}=N_0\left(\frac{E}{E_{0}}\right)^{-\alpha-\beta\cdot \textrm{log}\left(E/E_0\right)},
\label{eq:logparabola}
\end{equation}

\noindent where $N_0$ is the gamma-ray flux normalization, $E_0$ the pivot energy, $\alpha$ the gamma-ray spectral index and $\beta$ the curvature. Note this parametric form is reduced to a simple power law in the case of $\beta=0$. From this expression we can extract a useful parameter: the peak energy, $E_{\rm peak}$, i.e., the energy at which the energy spectrum $\left(E^2dN/dE\right)$ is maximum, by performing the consequent derivative, obtaining
$E_{\rm peak} = E_0\cdot e^{\frac{2-\alpha}{2\beta}}$, which represents a signature of different kind of emitting sources. In this work, we do not perform the spectral analysis of the 4FGL catalogue sources by ourselves. Instead, we use the parameters of the LP fit for each source published by the Fermi-LAT collaboration. \\ 
Similarly, we can predict the gamma-ray DM spectrum by means of Monte Carlo event generator softwares (see e.g., \citet{Cirelli:2010xx, Cembranos:2013cfa}). In fact, WIMPs annihilate in different SM channels, whose hadronization and decay processes generate spectra that are footprints of both the annihilation channel and the energy of the event, i.e. a signature of the DM candidate.
In \citep{Coronado-Blazquez:2019pny}, the authors introduced the DM in the $\beta-E_{\rm peak}$ parameter space (i.e. the $\beta$-plot) by fitting the DM gamma-ray spectrum, given by \citep{Cirelli:2010xx}, with the same LP functional form (Eq. \ref{eq:logparabola}). While in \citep{Coronado-Blazquez:2019pny} only pure annihilation channels $(B_r=1)$ were studied, we now consider more general two-channel linear combinations, of the form

\begin{equation}
\frac{dN}{dE}=B_r \left(\frac{dN}{dE}\right)_{C_1}+(1-B_r)\left(\frac{dN}{dE}\right)_{C_2}\,,
\end{equation}

\noindent where $C_1$ and $C_2$ are the two considered channels. We perform all possible combinations considering 10 branching ratios from 0 to 1 with a 0.1 step, for the annihilation channels $b\bar{b}, c\bar{c}, t\bar{t}, \tau^+\tau^-, e^+e^-, \mu^+\mu^-, W^+W^-, Z^0Z^0, \text{and}~hh$, and masses from 5 GeV\footnote{In some cases the lower mass is bounded by the mass of the particle itself, namely for the annihilation channels with $W^\pm$ ($m_{W^\pm}=80$ GeV), $Z^0$ ($m_{Z^0}=91$ GeV), $h$ ($m_h=125$ GeV) and $t/\bar{t}$ ($m_{t/\bar{t}}=173$ GeV).} to 10 TeV.\footnote{Although the spectra from \citep{Cirelli:2010xx} go to masses up to 100 TeV, the model-independent electroweak corrections used in these calculations are computed at leading order, while masses larger than $\sim$10 TeV, especially in leptonic channels, lack higher-order electroweak corrections not included in the tables, which may be relevant \citep{Cirelli:2010xx, Ciafaloni2011}. In any case, the LAT sensitivity quickly degrades at energies $\gtrsim 300$ GeV.} As we are agnostic to the underlying particle physics model that generates the annihilation, we consider all points as a ``DM cloud'' -- therefore being able to distinguish only between the astrophysical and DM scenarios, which is the ultimate goal of this paper. We generate a convenient number of DM points randomly distributed  within the boundaries of the DM parameter space.
The "DM-$\beta$"-plot is shown in Fig. \ref{fig:DM_beta_plot}.  In this plot, the orange points are astrophysical gamma-ray sources, the red points are detected unIDs and the magenta points are the DM sample. The overlap between DM and astrophysical sources is for light WIMPs and pulsars mainly, and especially in the case of hadronic channels such as $b\bar{b}$ and $c\bar{c}$, as expected \citep{fermi_dm_satellites_paper, Mirabal2013, Mirabal2016}. 
Nonetheless, a good portion of the region of the parameter space where the DM resides is radically different from the one where astrophysical sources lie. 

\section{Dark Matter systematic features}
\label{systematic}

In the previous section we have summarized and generalized the methodology of \citep{Coronado-Blazquez:2019pny} in order to introduce the WIMPs candidates in the $\beta$-plot parameter space, which allows us to train ML algorithms in order to distinguish and classify prospective DM-source candidates from astrophysical sources, only based on their gamma-ray spectra. \\ 
Nonetheless, such a description of the DM sample with only the two features of the $\beta$-plot, represents a limitation in the framework of ML. 
In fact, the collection of the unIDs sources we aim to classify includes a plethora of information - in terms of data or number of features - that are not considered in such a phenomenological DM data set.
Among other observational features that are not yet available for the DM sample, we will consider, on the one hand, the experimental systematic uncertainty $\beta_\text{rel}=\varepsilon_\beta/\beta$, of the curvature parameter $\beta$, and on the other hand the detection significance of the source, $\sigma_d$. Both quantities are of course inherent to both the identified sources and the unIDs. We explain below our procedure to artificially build such quantities for the DM sample. 

\subsection{Detection significance}

\begin{figure}
\begin{center}
\includegraphics[width=1\linewidth]{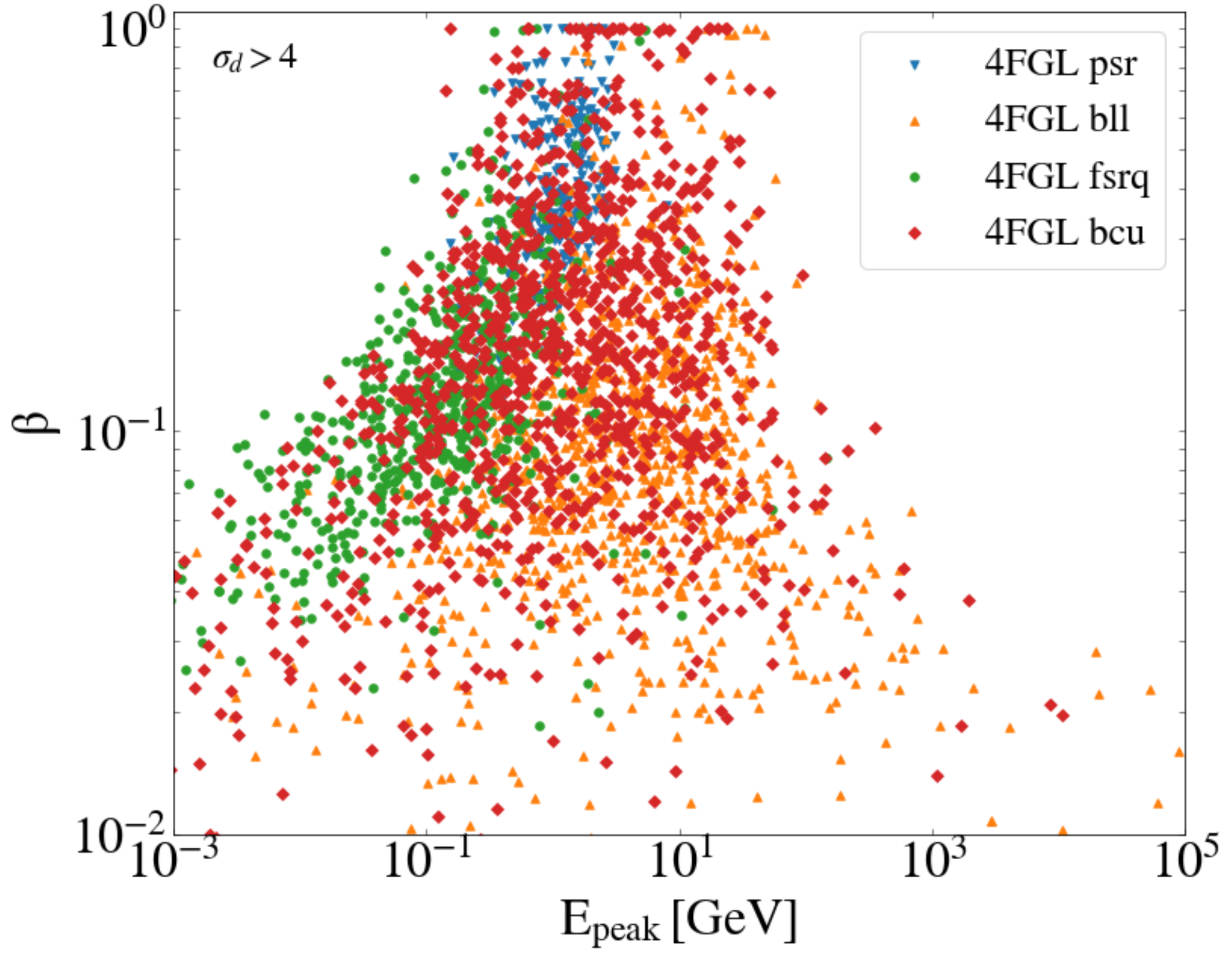}\\
\includegraphics[width=1\linewidth]{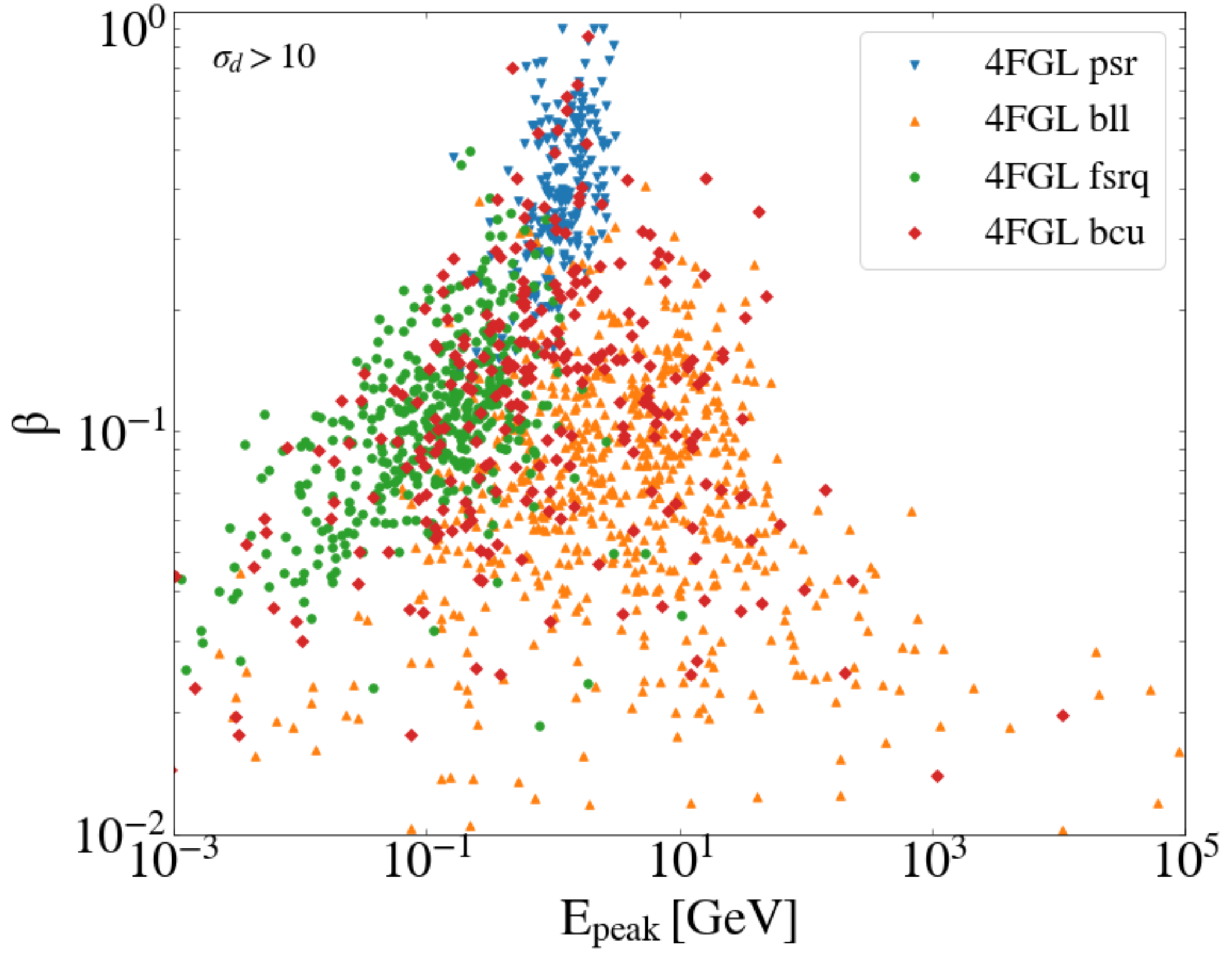}\\
\includegraphics[width=1\linewidth]{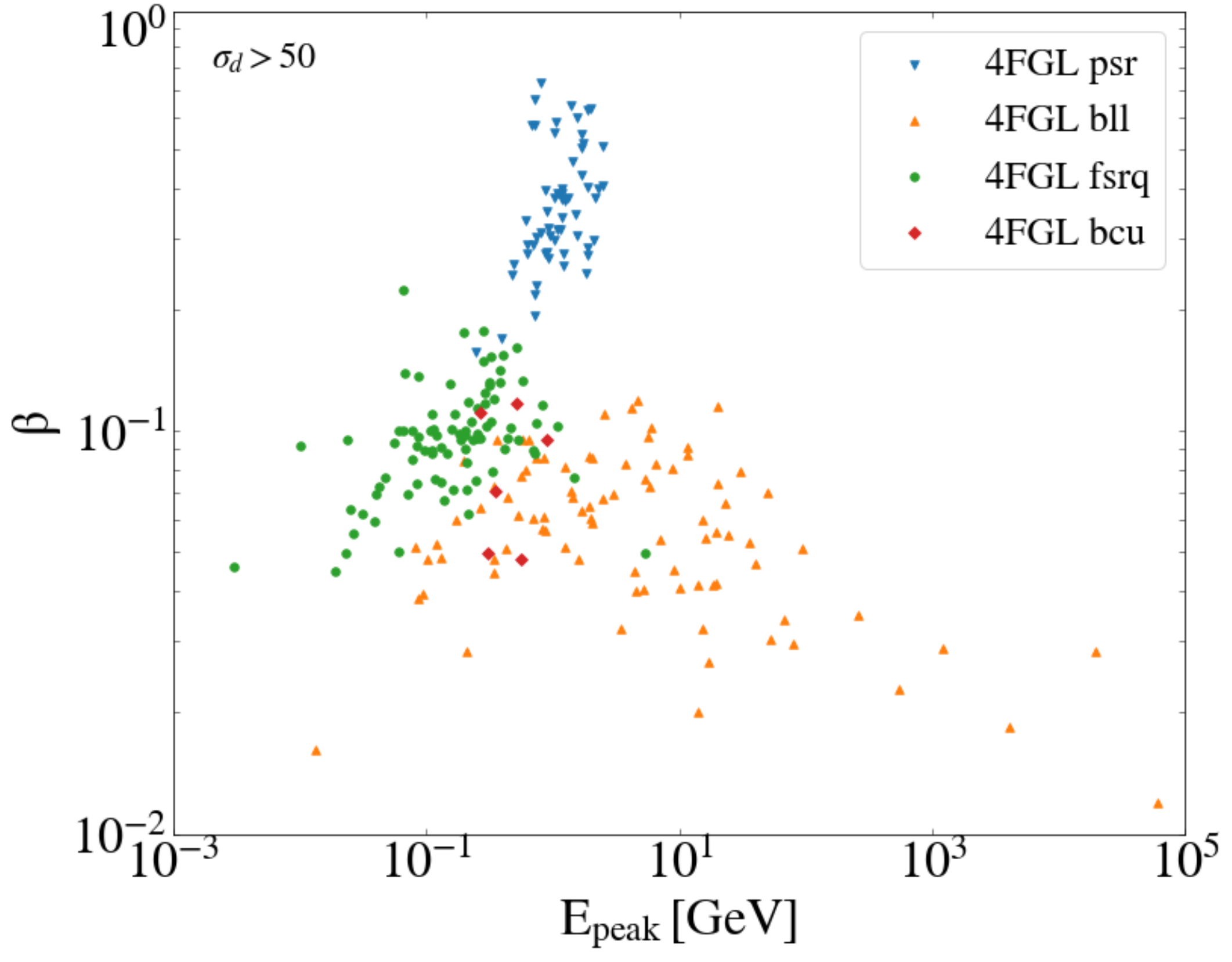}\\
\caption{\protect\footnotesize{The same as Fig. \ref{fig:DM_beta_plot} for 4FGL identified sources only and different cuts in detection significance $\sigma_{\rm d}$. Top panel: $\sigma_d>4$ (all sources). Middle panel: $\sigma_d>10$. Bottom panel: $\sigma_d>50$. Note the better separability of the classes as the cut is more stringent, at the cost of reducing the sample.}}
\label{fig:sigma_det}
\end{center}
\end{figure}

\begin{figure}
\centering
\includegraphics[width=1\linewidth]{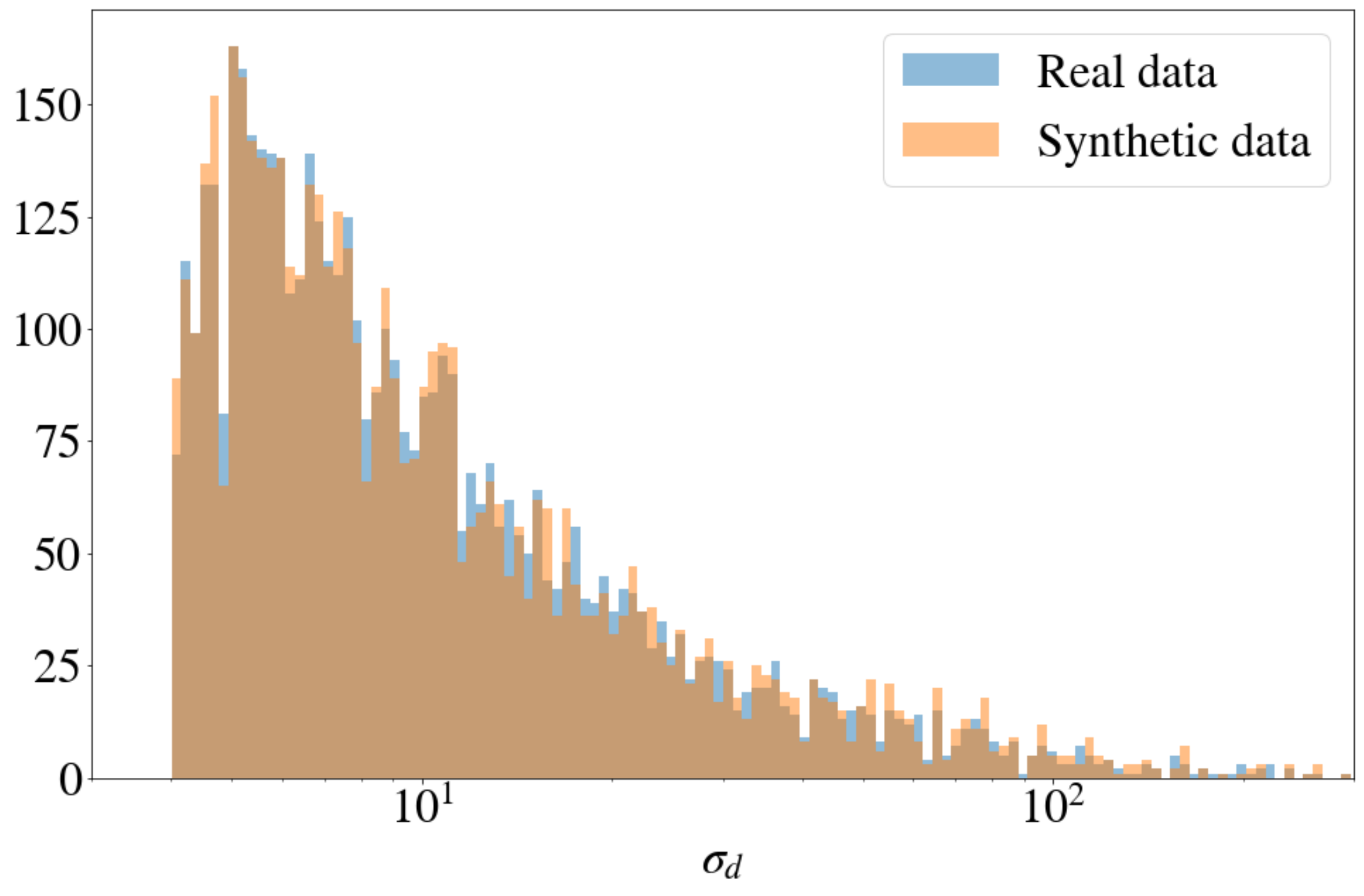}
\caption{\protect\footnotesize{Detection significance ($\sigma_{\rm d}$) distribution for the real unIDs data (blue) and systematic sampling for DM (orange), with a random noise similar to the one seen in the unIDs sample. The brown color only reflects the overlap of the previous two distributions.}}
\label{fig:sigmaTS_DM}
\end{figure}

First of all, it is phenomenologically interesting to note the different spread of the astrophysical classes in the $\beta-$plot. In Fig. \ref{fig:sigma_det} we show how the overlap between different astrophysical sources decreases by changing the cut applied on the detection significance, namely $\sigma_{\rm d}\geq 4,\, 10,\, 50$: the larger the significance, the smaller the overlap.\\
Generally speaking, the detection significance strictly depends on the data analysis. To analyze LAT data, the collaboration tools construct the likelihood that is applicable to the LAT data, and then use this likelihood to find the best fit model parameters. These parameters include the description of a source's spectrum, its position, and even whether it exists. 
Once that a template model of all the other sources in the source region is provided, the Test Statistic (TS) for adding an additional source at each gridpoint is calculated. The resulting significance is $\sim(TS)^{1/2}\sigma_d$, and thus TS=16-25 equivalent to $4-5\sigma_d$, is required for claiming the detection of any source in the 4FGL Fermi-LAT catalogue adopted in this work. The new source is characterized by a source intensity and spectral index \footnote{In a first approximation, the spectrum is assumed to be a power law}.
Hereafter, we will use the so-defined detection significance $\sigma_d$ as a \textit{systematic} feature of our classification problem. Note that, the DM data set has been created based on the WIMP phenomenology and the procedure outlined in Sec. \ref{sec:beta_plot}, and thus, it obviously lacks a detection significance, which is - by definition - an observational feature. In order to exploit the additional information coming from the distribution of the detection significance of the detected astrophysical sources, our idea is to build this variable as a fictitious feature of the DM class. The issue is not straightforward: in fact, the detection significance $\sigma_{\rm d}$ for the prospective DM sources would ultimately depend on many aspects, e.g. the WIMP mass, the SM annihilation channel, the Monte Carlo event generator software \citep{Cembranos:2013cfa}, the distance of the sources, the amount of DM in the source, as well as other hypotheses on the DM particle (see e.g. \citet{sym10110546}). If several DM subhalos were discovered, this class of DM sources would follow its own $\sigma_{\rm d}$ distribution ( see e.g. Sec. 3 of \citet{Gammaldi:2021zdm}).\\ 
As first hypothesis, we can assume that all the DM-source candidates are among the observed unIDs. We can therefore sample the unIDs $\sigma_{\rm d}$ distribution to generate mock data for DM with a random noise (Fig. \ref{fig:sigmaTS_DM}), such that the distribution is statistically the same but a single DM point in the DM-$\beta$ plot is assigned a random $\sigma_{\rm d}$. 
In this way, we associate to the theoretical DM sample a \textit{systematic} feature, which reflects systematics related to the adopted instrument, as shown in Fig. \ref{fig:histo}, first upper panel. 

\begin{figure}
\centering
\includegraphics[width=1.0\linewidth]{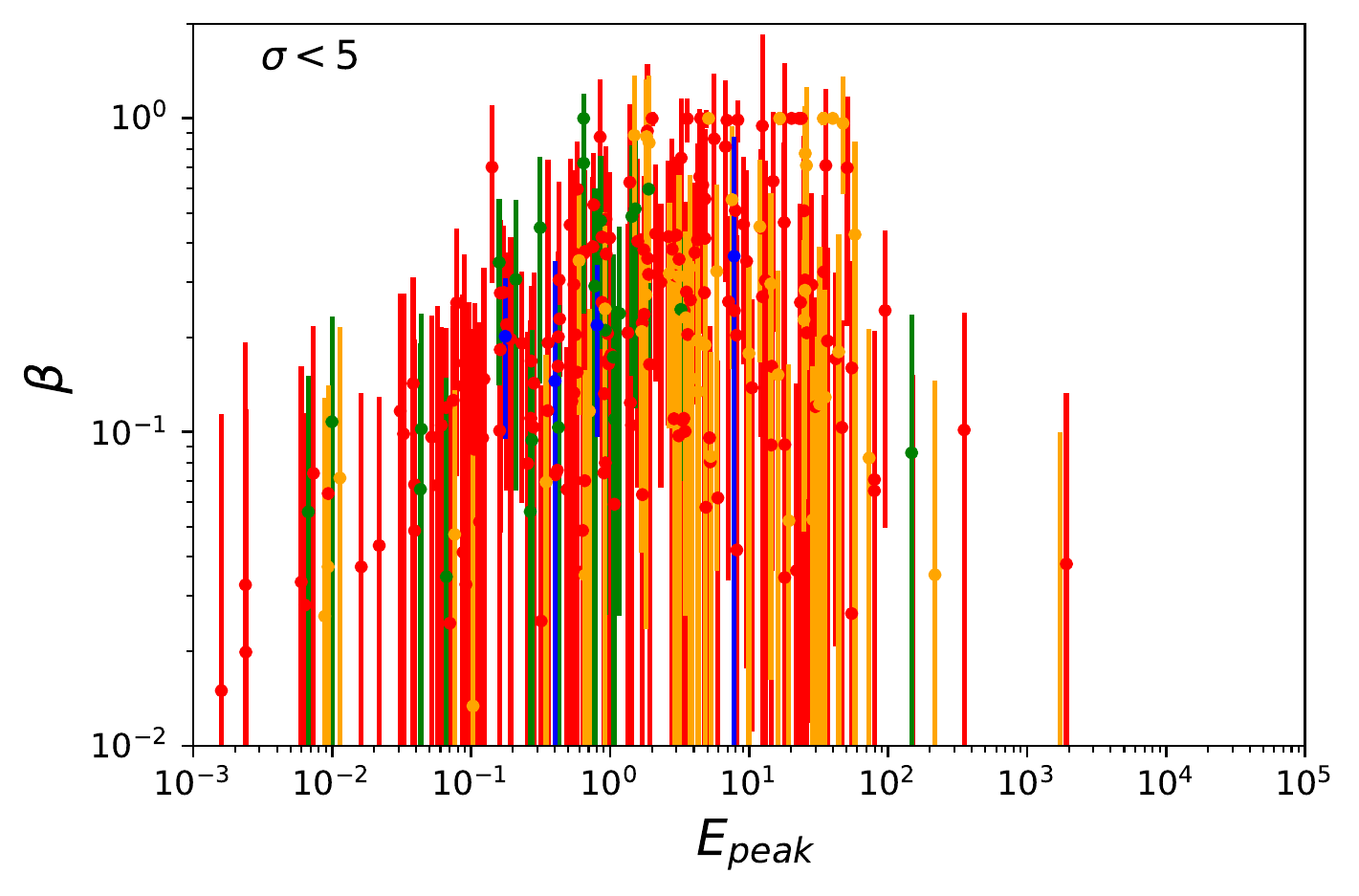}\\
\includegraphics[width=1.0\linewidth]{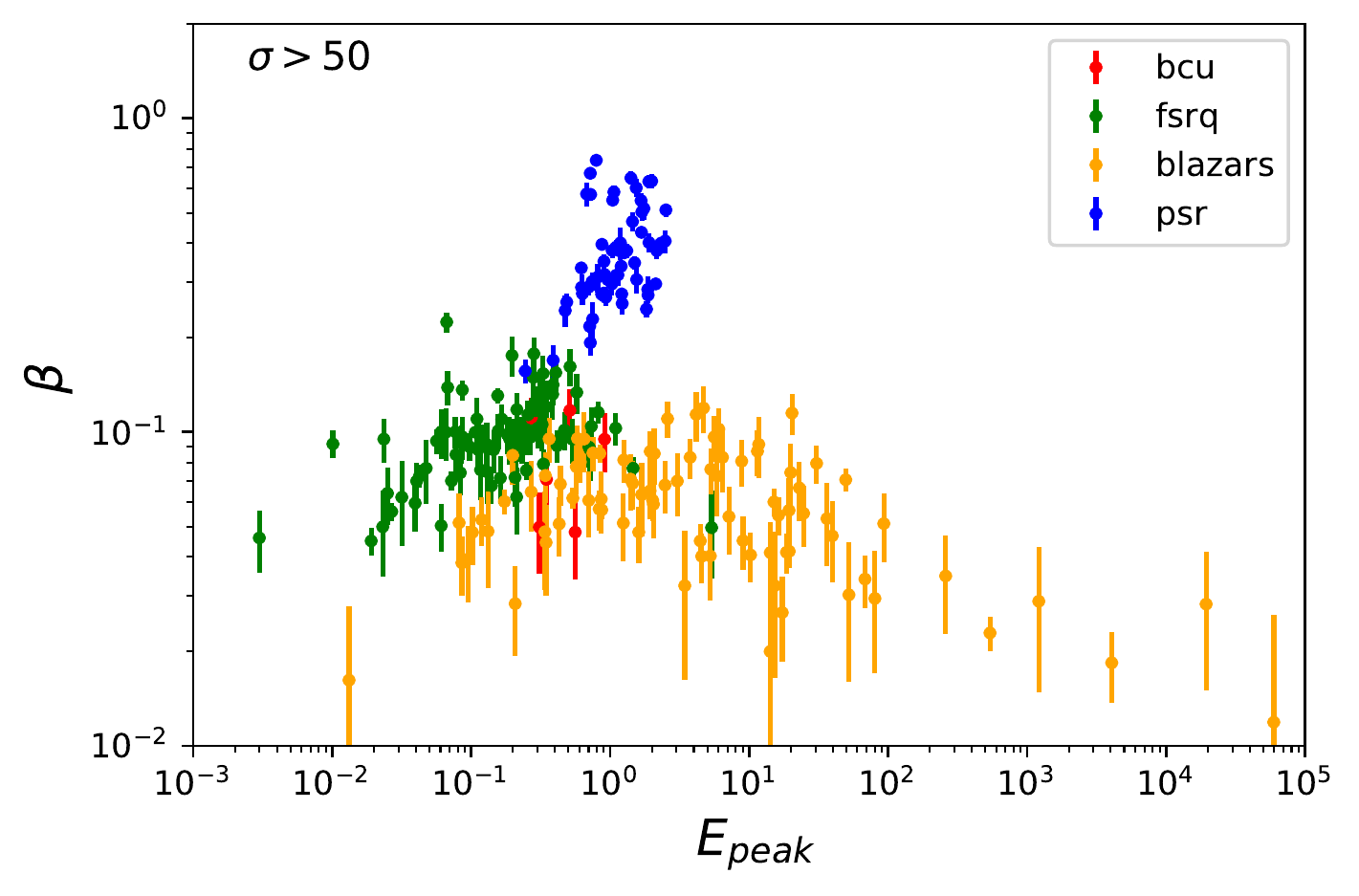}\\
\caption{\protect\footnotesize{Same as Fig. \ref{fig:sigma_det}, by including the uncertainty on $\beta$ for astrophysical data with $\sigma_d<5$ (upper panel) and $\sigma_d>50$ (lower panel). Let us stress as a lower detection significance corresponds to a worse characterization of the spectrum. Indeed, the classification is confused for data of lower $\sigma_d$ and improves for higher values of $\sigma_d$ also by eye. The color code in the legend is the same for both panels.}
}
\label{fig:sigma_det_uncertainty}
\end{figure}
\subsection{Uncertainty on \texorpdfstring{$\beta$}{beta}}
\begin{figure}
    \centering
     \includegraphics[width=\linewidth]{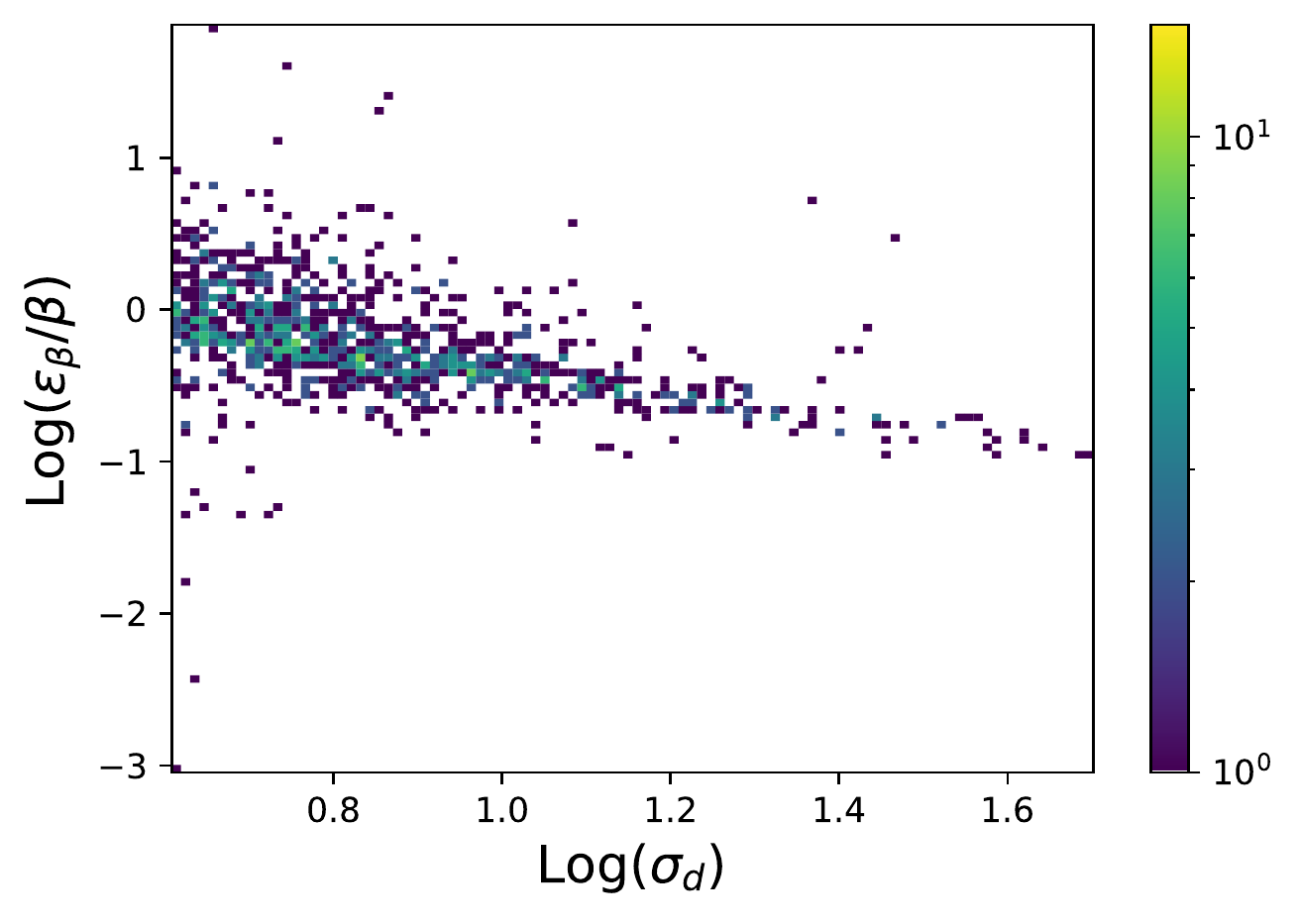}\\
    \includegraphics[width=\linewidth]{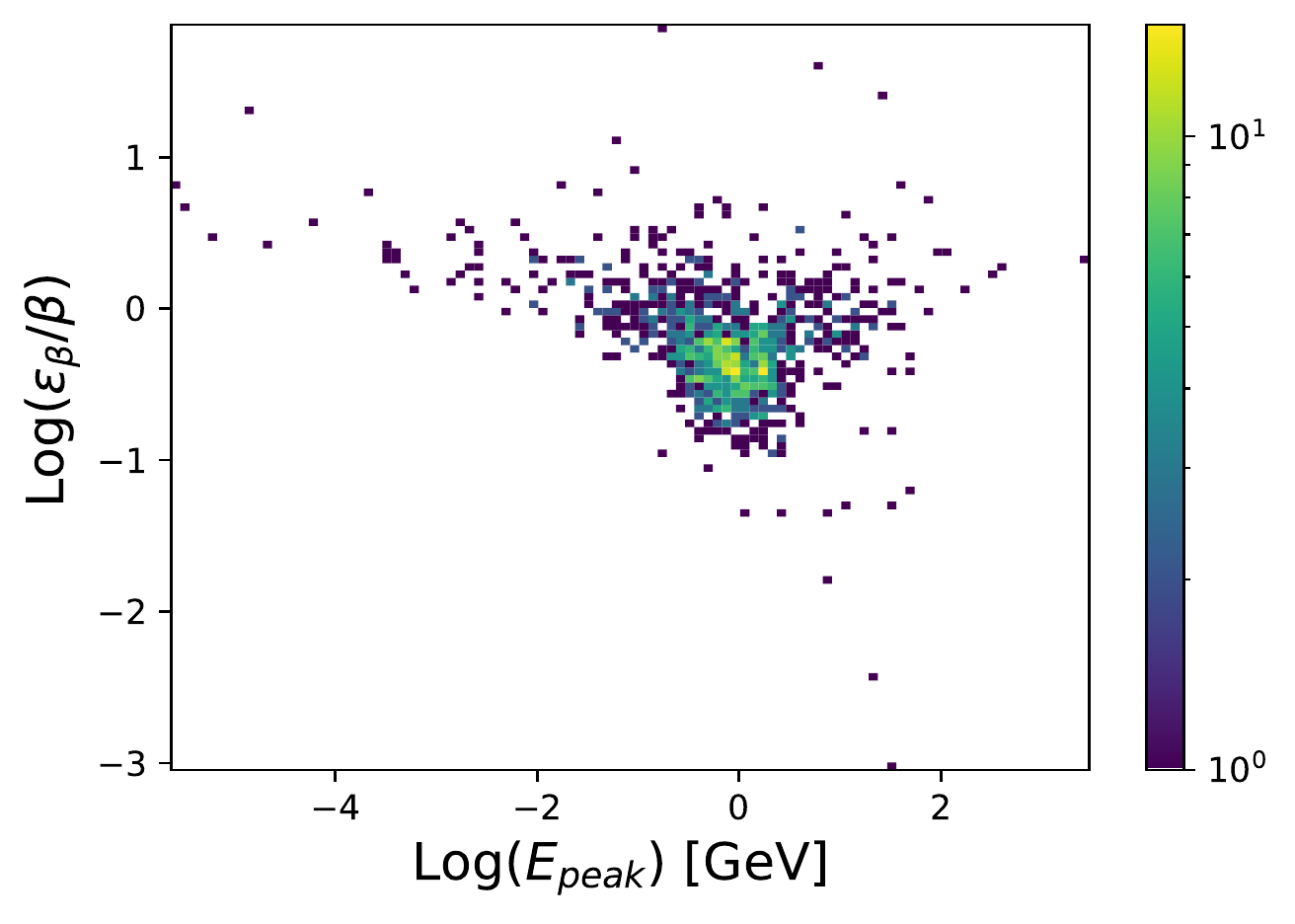}\\
    \caption{\protect\footnotesize{Upper panel: Relative error $\beta_\text{rel}=\varepsilon_\beta/\beta$ vs the detection significance $\sigma_\text{TS}$ for the 4FGL unIDs. Lower panel: Relative error, $\varepsilon_\beta/\beta$, vs. the $E_{\rm peak}$ for the 4FGL unIDs. Although there are sources with $E_{\rm peak}<0.3$ GeV, no DM point lie below this value, as the lightest WIMP mass considered is 5 GeV and the softest channel is $b\bar{b}$, which roughly peaks at $E_{\rm peak}\sim m_\chi/20$).  }}
    \label{fig:beta_err_unids}
\end{figure}

The next step of this analysis relies on the intuition that higher values of the detection significance $\sigma_{\rm d}$ correspond to better signal-to-noise ratio, i.e. to higher quality source spectra. When dealing with actual sources, the \textit{Fermi}-LAT standard analysis pipeline bins the spectral energy distribution (SED) and fits it with the corresponding parametric form (here, a log-parabola). The uncertainty in each bin will translate into an error in the parameters of the model. As a consequence, we expect a lower detection significance to correspond to a worse characterized spectrum. This translates into a higher uncertainty in the estimation of the spectral parameters $(E_\text{peak},\beta)$. We show qualitatively this property in Fig. \ref{fig:sigma_det_uncertainty} for astrophysical data with $\sigma_d<5$ (upper panel) and $\sigma_d>50$ (lower panel). 
The correlation between the relative error $\beta_{\rm rel}=\epsilon_\beta/\beta$ and the $\sigma_{\rm d}$ is shown in the upper panel of Fig. \ref{fig:beta_err_unids} for the unIDs population: clearly, the relative error $\beta_{\rm rel}=\epsilon_\beta/\beta$ decreases by increasing the detection significance $\sigma_{\rm d}$\footnote{Probably due to a poissonian statistical contribution to the total uncertainty.}, although they are not completely correlated.  In the lower panel of Fig. \ref{fig:beta_err_unids} we show the correlation between the relative error $\beta_{\rm rel}$ and $E_{\rm peak}$: in this case the correlation is slightly visible and it could be associated to the sensitivity of the instrument to different energy bins. In other words, there are more detected sources (with lower $\sigma_{\rm d}$ ) in the energy range where LAT is more sensitive ($\sim 1$ GeV), as expected. While partial correlations may exist among all the features considered in our analysis, a generic property of typical ML models is that they take them in into account automatically \citep{bishop}.  
This is exactly the kind of information that we aim to implicitly include via the analysis of these data within a ML approach instead of benchmark analyses. \\

As in the $\sigma_{\rm d}$ case, the DM data lack an observational $\beta_{\rm rel}$. Although a purely theoretical $\beta_{\rm rel}$ is given by the LP fitting of the simulated gamma-ray spectra expected by DM annihilation events \citep{Cirelli:2010xx}, we verified that such a theoretical error is below a few percent, and can be neglected with respect to the much larger systematic $\beta_{\rm rel}$ discussed so far. Such comparison is shown in the upper panel of Fig. \ref{fig:beta_err_DM}. The adopted distribution in shown in the lower panel of Fig. \ref{fig:beta_err_DM} and second panel of Fig. \ref{fig:histo}. 
\subsection{\texorpdfstring{$\beta_{\rm rel}$}{} as \textit{systematic} feature}
Analogously to the $\sigma_{\rm d}$ feature, we can sample the distribution of $\beta_{\rm rel}$ for the unIDs population to associate uncertainties on $\beta$ to the DM data. Here we consider a 2D sampling space, as the $\varepsilon_\beta$ depends also on the $E_{\rm peak}$. Indeed, from Fig. \ref{fig:beta_err_unids} (lower panel), one can see that there is a cluster at $E_{\rm peak}\sim0.5-2$ GeV, while for energies above $\sim 60$GeV the errors tend to be larger as the statistics of unIDs decreases. This is due to the LAT sensitivity, which reaches its maximum at 1-2 GeV.

\begin{figure}
\begin{center}
\includegraphics[width=0.46\textwidth]{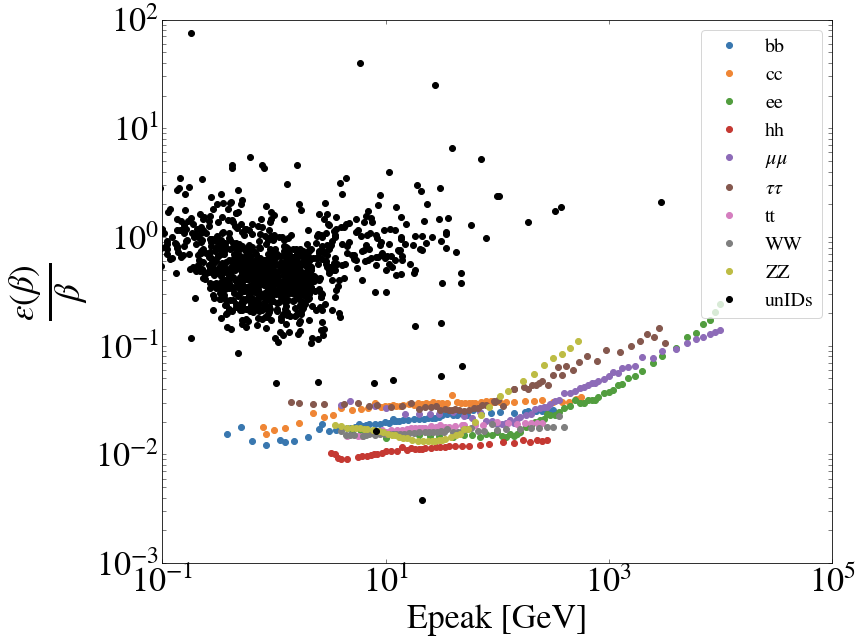}\\
\includegraphics[width=1.0\linewidth]{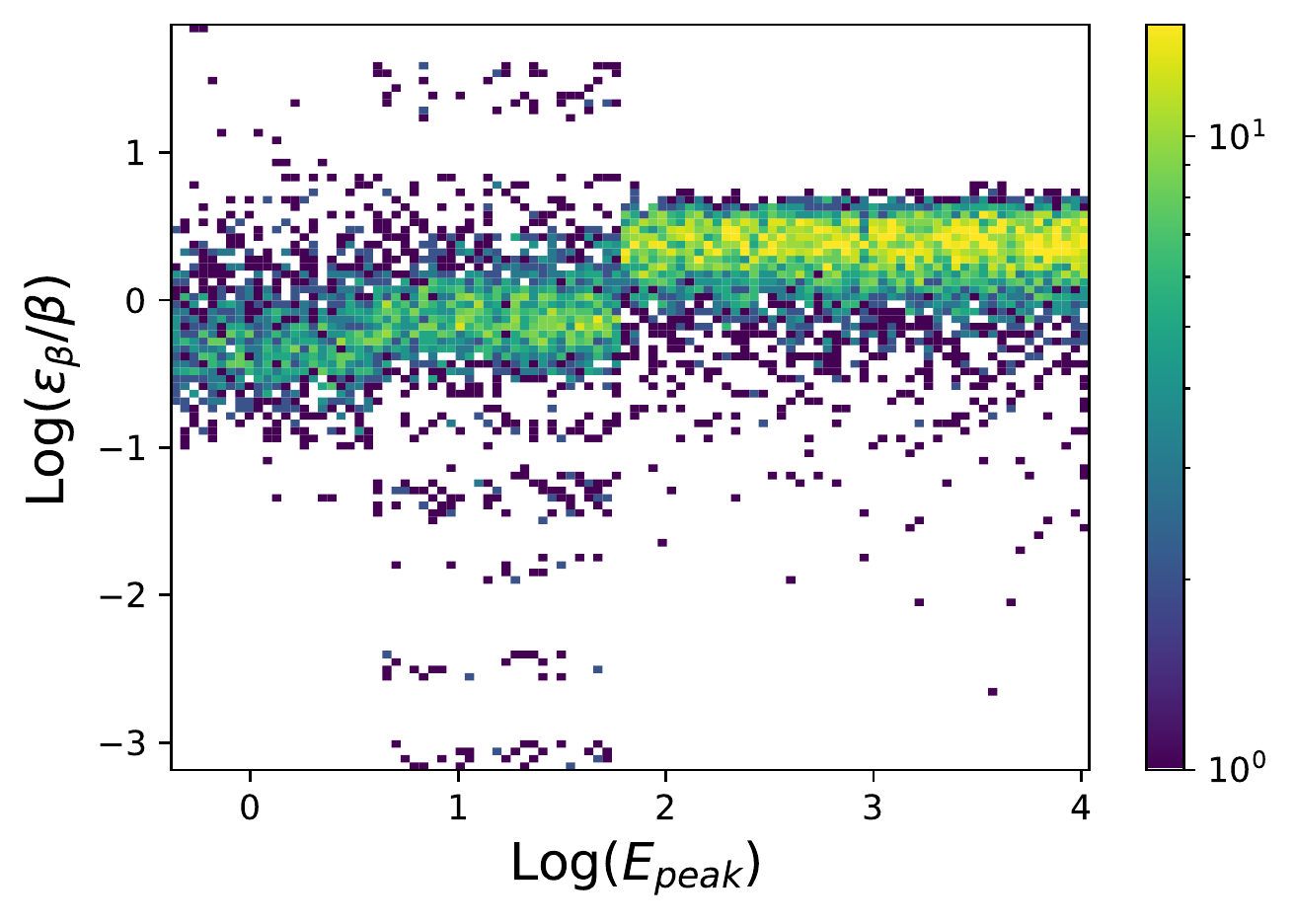}\\
\caption{\protect\footnotesize{Upper panel: experimental vs theoretical $\beta_{\rm rel}$. Lower panel: relative error, $\beta_{\rm rel}=\varepsilon_\beta/\beta$, vs. the $E_{\rm peak}$ for the DM data, sampled from the 4FGL unIDs. Three regimes are considered, $E_{\rm peak}<4$ GeV, $4<E_{\rm peak}<60$ GeV and $E_{\rm peak}>60$ GeV.}}
\label{fig:beta_err_DM}
\end{center}
\end{figure}

\begin{figure}
\begin{center}
\includegraphics[width=0.40\textwidth]{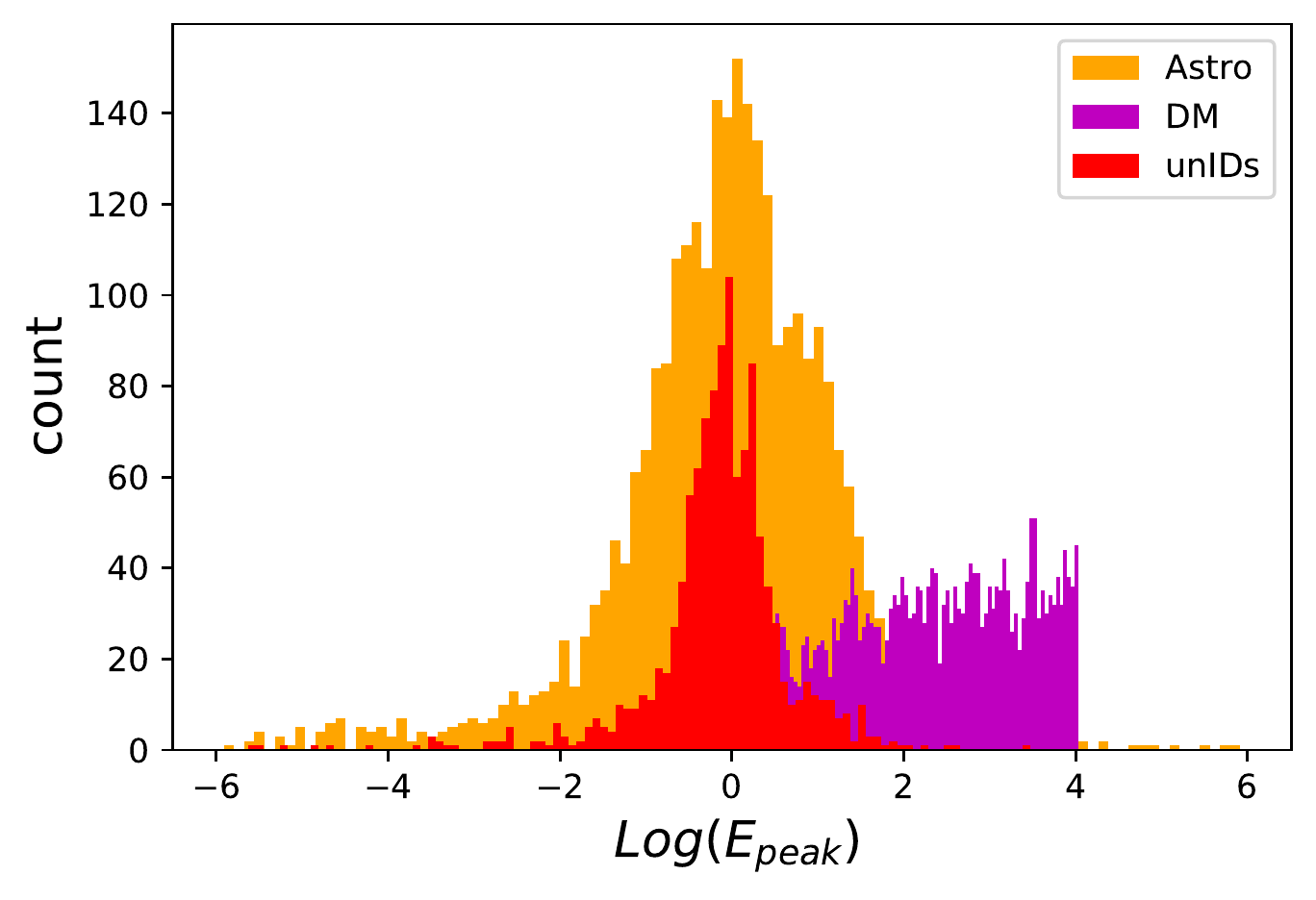}\\
\includegraphics[width=0.40\textwidth]{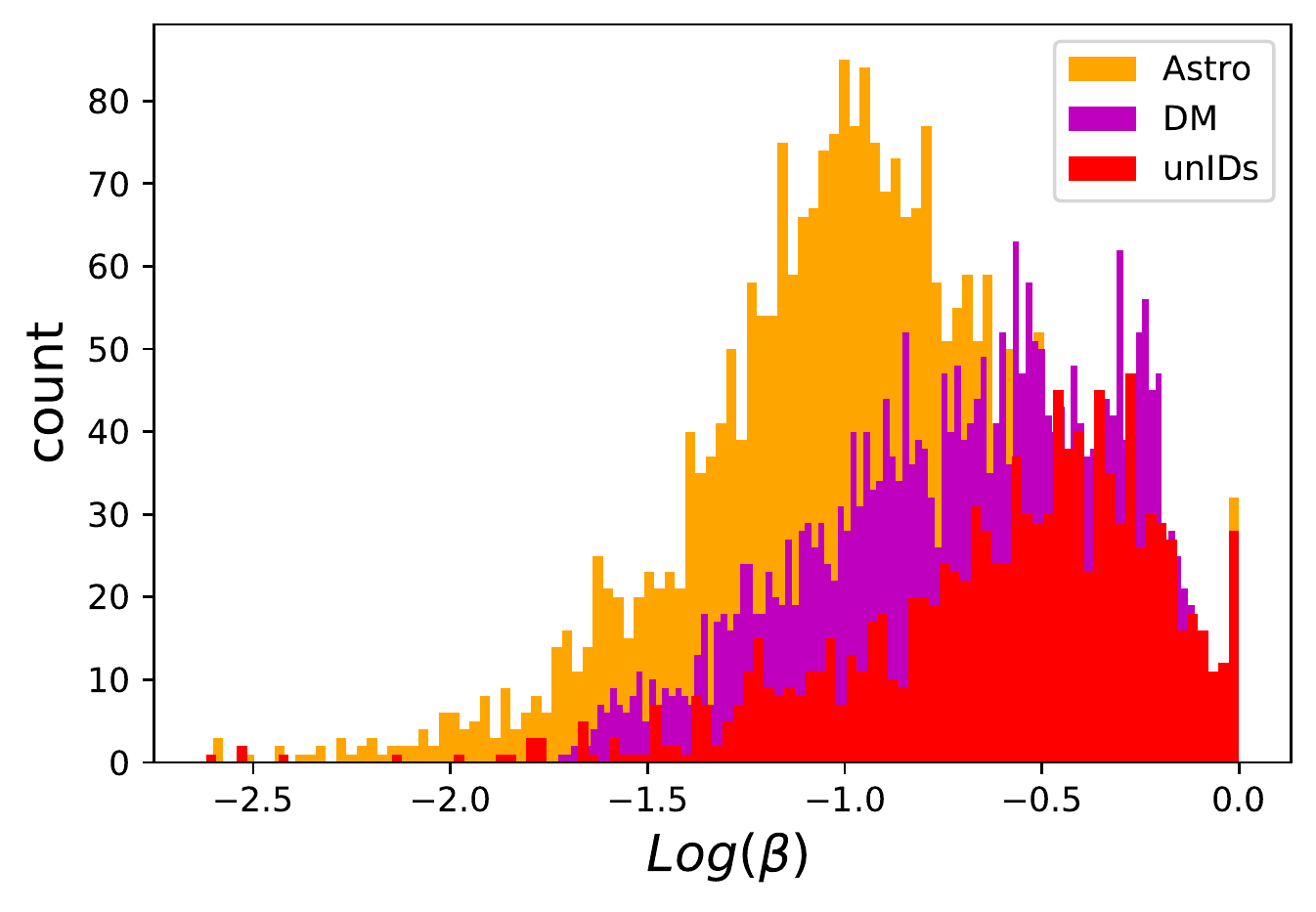}\\
\includegraphics[width=0.40\textwidth]{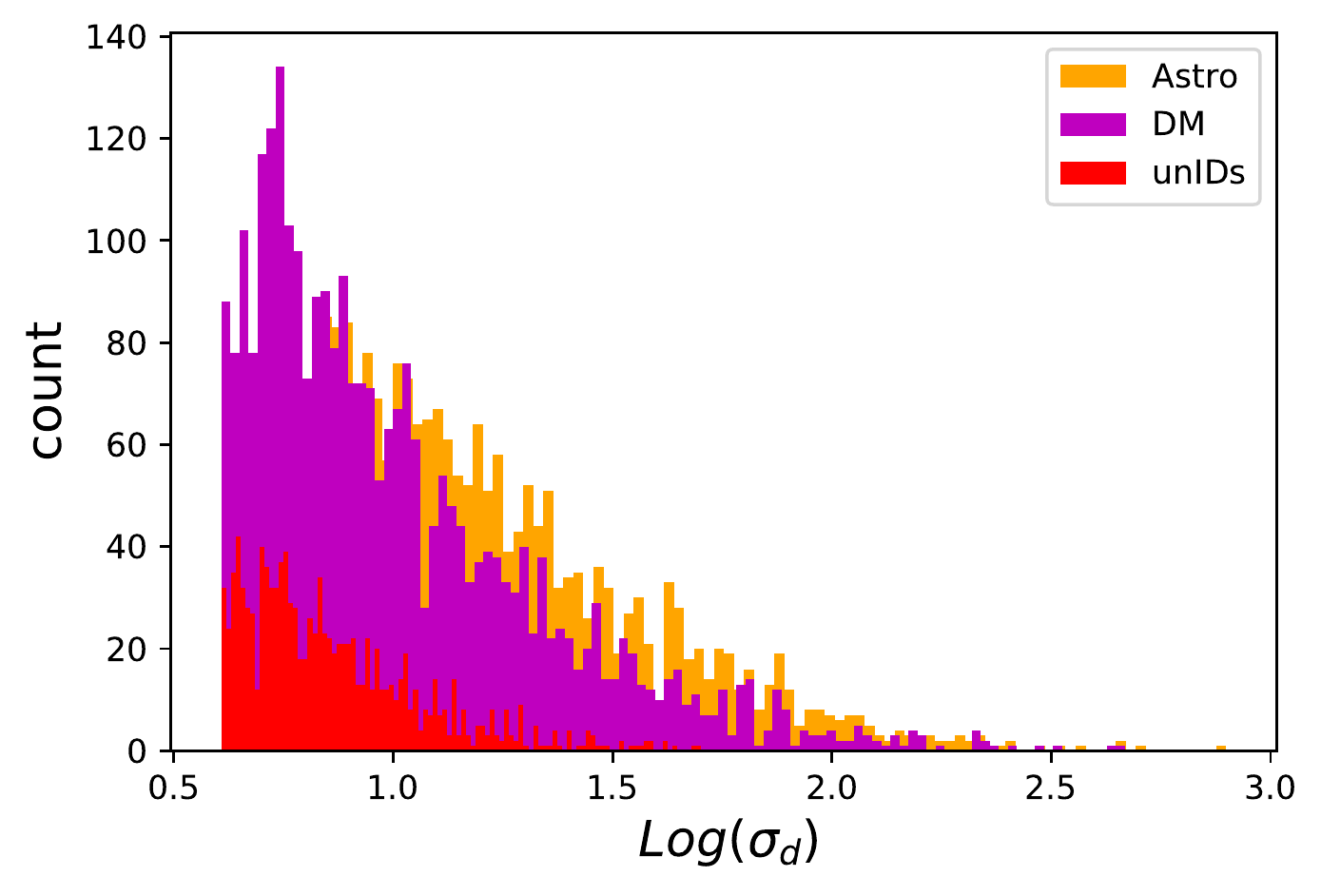}\\
\includegraphics[width=0.40\textwidth]{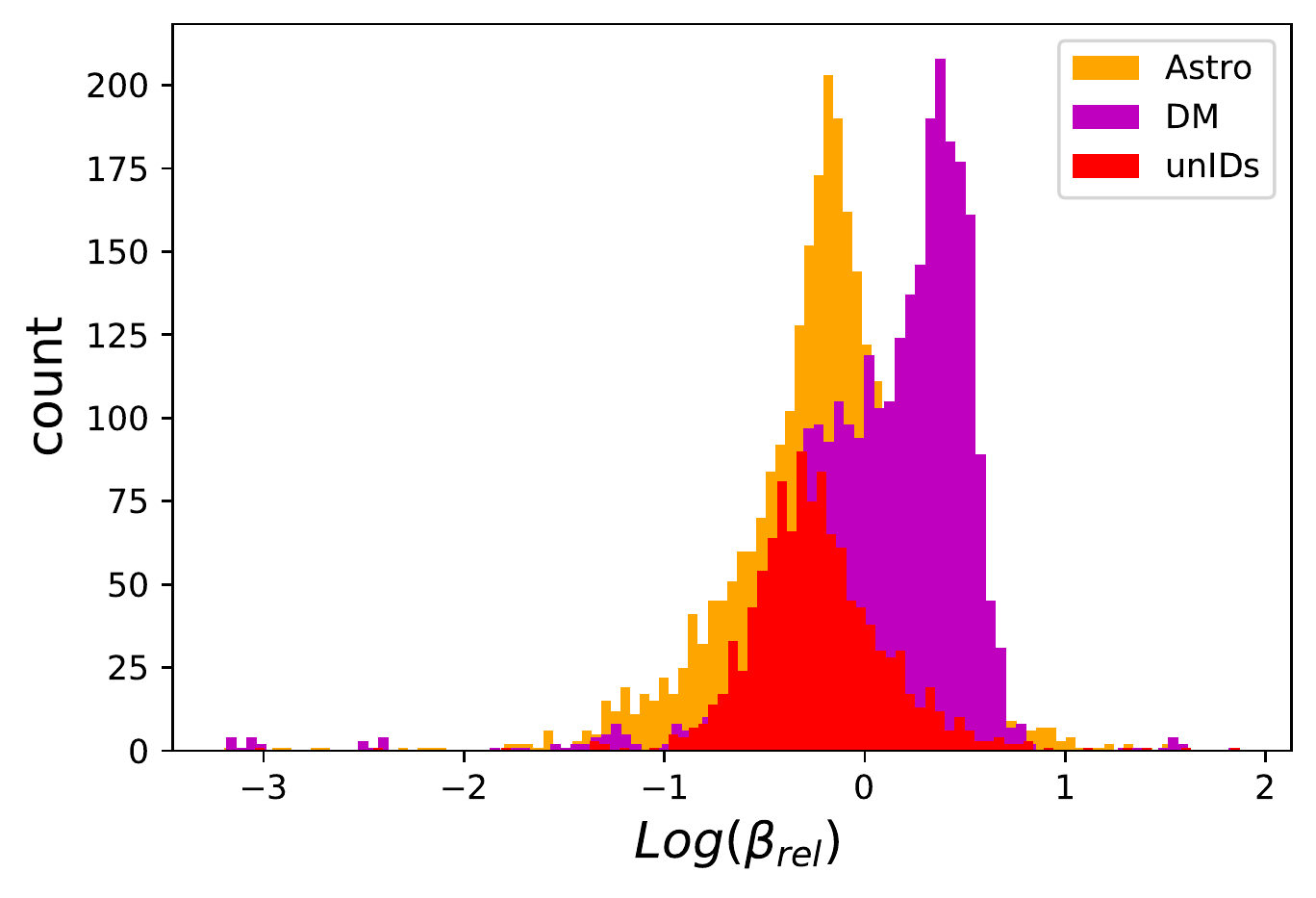}\\
\caption{\protect\footnotesize{Histograms of the four features of the balanced data\protect\footnote{balanced data means that the number of data in different classes, here astro and DM, is kept of the same order.} adopted. From upper to lower panel: emission energy  $E_{\rm peak}$, curvature of the spectra $\beta$, detection significance $\sigma_{\rm d}$, relative error on $\beta$. In each panel we show the histograms for the classified astrophysical sources (orange), unIDs (red) and DM data set (magenta).}}
\label{fig:histo}
\end{center}
\end{figure}

In order to assign the DM $\beta_{\rm rel}$ systematic values, we will divide the distribution in three bins, $E_{\rm peak}<4$ GeV, $4<E_{\rm peak}<60$ GeV and $E_{\rm peak}>60$ GeV. According to Fig. \ref{fig:beta_err_unids}, these boundaries approximately reflect three different regimes in the data, with a cluster of objects in the first one, a more spread distribution in the second and a third one where no source with $\varepsilon_\beta/\beta<1$ is found. Sampling directly the unbinned distribution would lead to under-estimated errors for the highest $E_{\rm peak}$ values. As the points with $E_{\rm peak}>60$ GeV are very scarce (just 7), we introduce a random Gaussian noise to avoid discreteness in the distribution. For consistency, we do the same for the other two bins.
The result of this sampling is shown in the lower panel of Fig. \ref{fig:beta_err_DM}. The last bin is the most populated one, as it is the one which contains more DM points (mostly due to hard channels such as $e^+e^-$ and $\mu^+\mu^-$, which peak at $E_{\rm peak}=m_\chi$). 
The pronounced step visible in the figure at $E_{\rm peak}=60$ GeV is simply caused by the binning choice, and can also be seen in the original unIDs distribution of Fig. \ref{fig:beta_err_unids}. \\
With these two samplings of $\sigma_{\rm d}$ and $\beta_{\rm rel}$, we have generated two equivalent datasets consisting on $\{E_{\rm peak}, \beta, \beta_{\rm rel}, \sigma_d\}$, both for the 4FGL catalog and DM. The \textit{systematic} distribution for $\sigma_d$ and $\beta_{\rm rel}$ created for the theoretical DM sample (magenta histograms) are shown in the two lower panels of Fig. \ref{fig:histo} - as well as the distributions for the observed data, i.e. astrophysical sources (orange histograms) and the unIDs (red histograms). Note also that, although we adopt in our analysis the LP parameters given by the Fermi-LAT collaboration for the full catalogue of detected sources, we take into account the possibility that any Fermi-LAT source could be not well fitted with a LP by including the $\beta_{\rm rel}$ uncertainty in the 4F analysis.  in Appendix \ref{DM_beta_uncertainty} and Fig. \ref{fig:beta_plot_errors_20sigma} we show an example of the $\beta-$plot with the astrophysical and DM data set including the systematic uncertainty on $\beta$. 

\section{Methodology}
\label{Methodology}

\subsection{Classification algorithms}

We interpret the problem as a standard binary classification task in ML. Data consists of ${\cal D}=\{{\bf x}_i, t_i\}$, being $i=1,..,N$, where ${\bf x} = \{E_{\rm peak}, \beta, \beta_{\rm rel}, \sigma_d\}$ is the multivariate input and a label $t_i=\{0,1\}$, corresponding to astro or DM class, respectively. 

We study the performance of several ML models, coming from different approaches and having different levels of expressiveness. They are briefly specified next:
\newline\noindent
\begin{itemize}
\item Probabilistic discriminative models.
In order to estimate the expected value $y_i$ of the label $t_i$, we start with the simplest classifier: the Logistic Regression (LR) model. 
Secondly, we use a fully-connected feed-forward neural network (NN) with one hidden layer. See Appendix \ref{sec:repro} for further technical details of the implementation. Both models aim at estimating the probability $p(C_k | {\bf x})$ of class $C_k$ given the input ${\bf x}$, which will depend on some parameters to be optimised. \\

\item Generative model. We also consider the Na\"ive Bayes (NB) classifier, where the likelihood $p({\bf x}_i|C_k)$ of point $i$ given a class $C_k$ is taken as a multivariate Gaussian distribution, with diagonal covariance matrix. This is a common benchmark model, since even if not requiring numerical optimisation, it is typically giving reasonably good results also in real-world datasets. We have used our own {\tt python} implementation of this model.
\\

\item Non-parametric model. Finally, we consider a specific Gaussian Process classifier \footnote{See \cite{rasmussen:williams:2006} for the classical textbook about Gaussian Processes.}, namely Noisy Input Multi-class Gaussian Process (NIMGP)\footnote{Here modified to solve a binary classification problem.} (\citet{2020arXiv200110523V}), which was constructed in such a way to incorporate the uncertainties of the input variables, either given explicitly (say, from the experiment, as it is our case at hand), or to be learned by the model itself. While more details are given in Appendix \ref{NPMGP}, in short here the idea of such model is to assume that every observation ${\bf x}_i$ is a noisy instance of the true value (call it ${\bf\tilde x}_i$), following a Gaussian distribution.       
\end{itemize}

\subsection{Setups}
\label{setups}
In this work we consider different setups for the classification task, with increasing level of complexity either from the dataset itself as well as from the modelling part.
They are described below:
\newline\noindent
\begin{itemize}

\item ``{\bf 2F}''. A dataset with only two features: $E_{\rm peak}$ and $\beta$. This is the minimum setup, not requiring the construction of additional variables beyond those coming from the fit of the spectrum for both astrophysical sources and DM. This setup does not take into account the uncertainty on $\beta$ when doing the classification.\\

\item ``{\bf 3F-A}''. A first strategy for taking into account the uncertainty on $\beta$ is to follow an heuristic, by which the original dataset is artificially augmented, assuming that every observation $\beta_i$ follows a Gaussian, whose mean is given by the precisely observed value, and the standard deviation is the reported uncertainty on $\beta_i$. Then we augment the dataset by taking each original point and sampling 60 times from it. coincides An augmented dataset containing three features:  $E_{\rm peak},~\sigma_{\rm d}$ and $\beta_{\rm sampled}$. More details can be found in the appendix  Sec. \ref{sec:augmented_sample} and Fig. \ref{fig:histo_sampled}.\\ 

\item ``{\bf 3F-B}''. The second strategy for incorporating the uncertainties in $\beta$ is inspired in a recent work by \cite{2020arXiv200110523V}, as commented above and explained in more detail in the appendix. This is arguably the most formal procedure for taking into account the input uncertainties, among all the setups we consider here. The dataset here contains the three same features as above, i.e. $E_{\rm peak},~\sigma_{\rm d}$ and $\beta$. However, now the uncertainties of $\beta$ are just included in the statistical model. Concretely, this setup will concern exclusively the NIMGP model mentioned above.\\   

\item ``{\bf 4F}''. The last strategy for taking into account the uncertainties of $beta$, is to include them as a separate feature (input variable). This is a priori reasonable, since we have checked that there is only a minor correlation between $\beta$ and $\epsilon_\beta$.\footnote{Actually, the Pearson correlation coefficient being igual to 0.4.}. The dataset here contains four features:  $E_{\rm peak},~\sigma_{\rm d},~\beta$ and $\beta_{\rm rel}$. Note that, in the case of the DM class, both $\sigma_{\rm d}$ and $\beta_{\rm rel}$ have been constructed out of the unIDs population, as discussed in Sec. \ref{systematic}. \\

\end{itemize}

\label{sec:results}

\subsection{Data Pre-processing}
\label{data}

We pre-process the data as follows:

\begin{itemize}
  \item we apply a cut on $10^{-3}\leq E_{\rm peak} \leq 10^{6} $, which is a reliable range of energy due to the Fermi-LAT sensitivity;
    \item we create the DM data sample in order to have a balanced data set, i.e.  the same number of astrophysical (hereafter, astro) and DM data;
    \item we work in log-space, due to the broad range of values for each feature;  
    \item we standardize data (see e.g. \citet{SHANKER1996385, bishop}), i.e. each feature is transformed to have zero mean and unit variance. For each classification run, the standardization is done with respect to the training data set and testing data set, independently. The unIDs sample has been also standardized. 
\end{itemize}

\section{Astro-vs-DM classification results}
\label{sec:class_acc}

In the following we show the performance of different combinations of the four ML algorithms and setups previously introduced. \\
\\
Firstly, we consider three models: the LR, NN and NB models for three of the different setups described in Sec. \ref{setups} (the 2F, 3F-A and 4F) and the GP for the 3F-B setup. The results are shown in Table \ref{tab_OA}, where the columns show the overall classification accuracy (OA), the True Negative (TN) rate, and the True Positive (TP) rate, respectively, while noting that ``negative'' here refers to the astro class, while ``positive'' refers to the DM class. The False negative (positive) rate can be simply obtained as $100\%- \text{TP (TN)}$, being these values normalized over the true.
The reported value and quoted uncertainty correspond to the mean and standard deviation of the OA, TN/P rate obtained after 100 splits (see Appendix \ref{sec:Nfolds} for further details). The precision $P=TP/(TP + FP)$ and the False Discovery Rate $FDR=FP / (TP + FP)$ may be also deduced from the table: for the NN-4F we have $P=0.94\pm 0.02$ and $FDR=0.06\pm 0.02$.
We find out that all the classifiers improve their OA, TN and TP from the 2F to the 4F setup, by including the systematic features. On the other hand, the accuracy decreases for the ``3F-A'' configuration, for all the three classifiers. The reason for this is simply that, in the augmented dataset, the two classes will necessarily overlap more, quantitatively depending on the quoted uncertainty for $\beta$. We conclude that among the two strategies considered so far for taking into account $\beta_{\rm rel}$, the 4F setup gives better performance\footnote{We have verified that a further setup with 3F ($E_{\rm peak}, \beta, \beta_{rel}$) returns OA$=88.6\%\pm 0.8\%$, TN$=87.9\% \pm 2.8\%$, TP$=89.2\% \pm 2.8\%$, indeed worst than the NN in the 4F setup.}. 
We get OA$= 93.3\% \pm 0.7\%$ and we can correctly classify $94.7\% \pm 1.7\%$ of astrophysical sources. Intuitively, we give more importance to the correct classification of already well-known astrophysical sources (TN) than to the one of prospective DM sources (TP), i.e. to a second level, our best classifier will be the one that maximizes not only the OA, but also the TN percentage. 

\begin{table}
\resizebox{\columnwidth}{!}{
\begin{tabular}{|c|c|c|c|}
\hline 
\hline
$~$ & OA($\%$) & TN ($\%$)  & TP ($\%$)   \\ 
\hline
\hline
\hline
LR &&&\\
\hline
\hline
2F & $84.9\pm 0.8$   & $85.4\pm1.5$
 & $84.4\pm1.4$   \\
 \hline
3F-A & $83.0\pm 0.1$ & $85.0\pm0.2$ & $81.0\pm. 0.2$ \\
\hline
4F & $86.0\pm0.9$ & $86.7\pm1.5$  & $85.2\pm 1.3$\\
\hline
\hline\hline
NN  &&& \\
\hline
\hline
2F   & $86.2\pm0.8$   & $86.1\pm3.0$  & $86.4\pm3.4$     \\
\hline
3F-A   & $85.0\pm 0.2$ & $87.9\pm1.8$ & $82.3\pm. 1.8$  \\
\hline
4F   & ${\bf 93.3\pm0.7}$   & ${\bf 94.7\pm1.7}$ & ${\bf 91.8\pm 1.5}$   \\
\hline
\hline
\hline
NB  &&& \\
\hline
\hline
2F  & $82.4\pm 1.5$   
& $83.9\pm 1.9$  
& $80.5\pm 2.5$   \\
\hline
3F-A & $82.5\pm 0.3$ 
   & $83.7\pm 0.4$  
   & $81.6\pm 0.3$  \\
\hline
4F & $83.5\pm 1.0$ 
   & $86.2\pm 1.2$  
   & $81.7\pm 1.2$  \\
\hline
\hline\hline
GP &&& \\
\hline\hline
3F-B  & $88.1\pm 0.2$ & $89.6\pm 0.3$ & $84.9\pm 0.2$ \\
\hline
\hline
\end{tabular}}
\caption{\protect\footnotesize{Performances of the three different ML models LR, NN, NB in the 2F, 3F-A and 4F setups compared with the GP in the 3F-B setup, as described in \ref{sec:class_acc}. See text for details. We highlight in bold face the results of the configuration giving the best performance.}}
\label{tab_OA} 
\end{table}

Finally, Table \ref{tab_OA} also shows the result of the GP model (specifically, the NIMGP implementation, see Appendix \ref{NPMGP}) using the 3F-B configuration. Even if this is the more complex model considered (in number of the free parameters of the model to be optimized), we see that its performance concerning classification accuracy is smaller than for the NN model with the 4F setup. This is not surprising: indeed it is common that large Bayesian models may show an overall performance which is not higher than flexible models in a frequentist approach (as the NN). Instead, the advantage of using Bayesian inference comes mainly from its capability of providing estimates of prediction uncertainties, accounting for both statistical (a.k.a aleatoric) and modelling  (a.k.a. epistemic) uncertainties. 
The way to do it is via the predictive distribution $p(y_*|{\bf x}_*,{\cal D})$, for the class $y_*$ at a new input ${\bf x}_*$, given the already observed (training) data ${\cal D}$. This is an intrinsically Bayesian quantity, and does not have counterpart with the frequentist implementation of the NN we have adopted in this work.

\section{UnIDs classification}
\label{sec:unids_class}

\begin{figure}
    \centering
   \includegraphics[width=\linewidth]{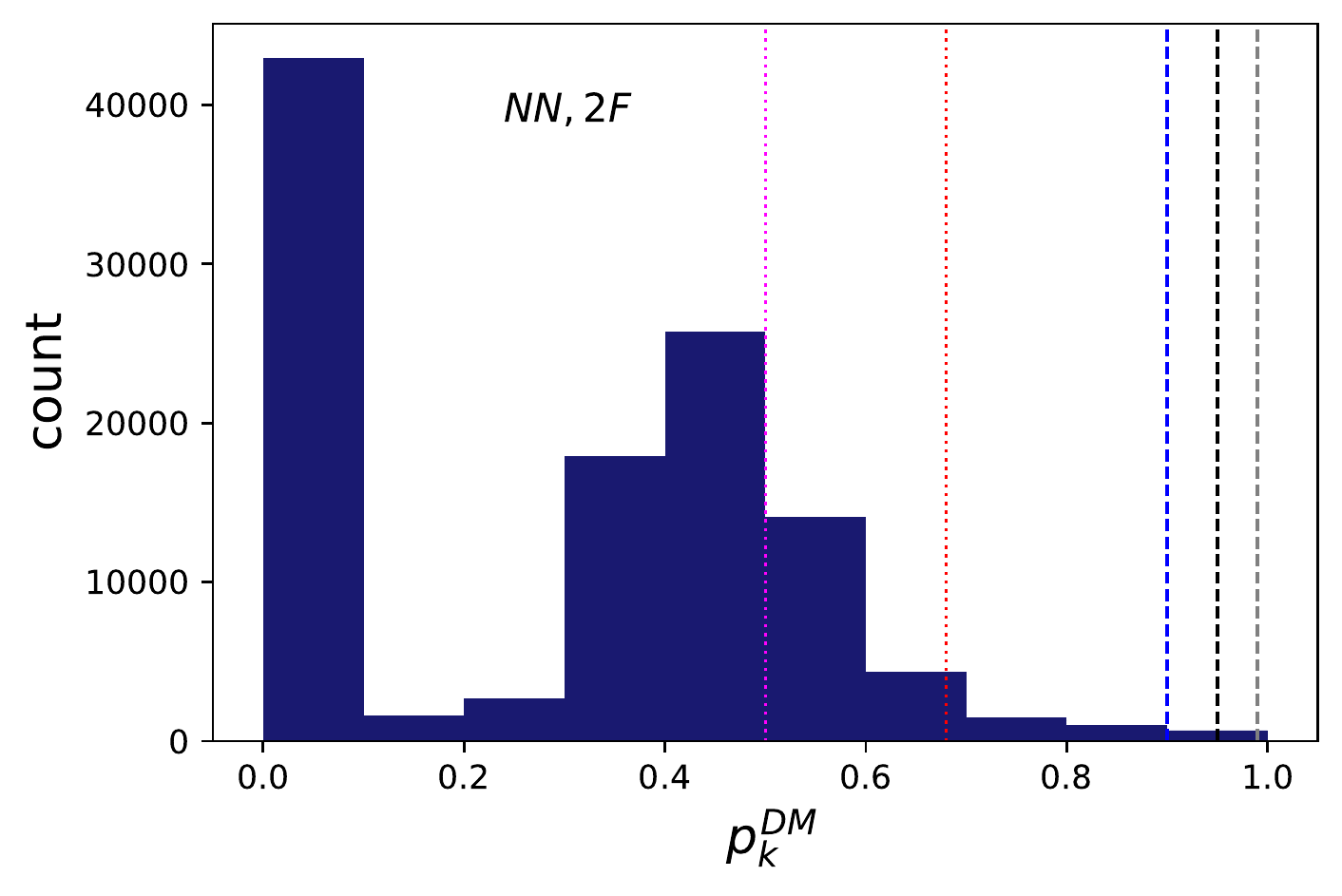}
   \includegraphics[width=\linewidth]{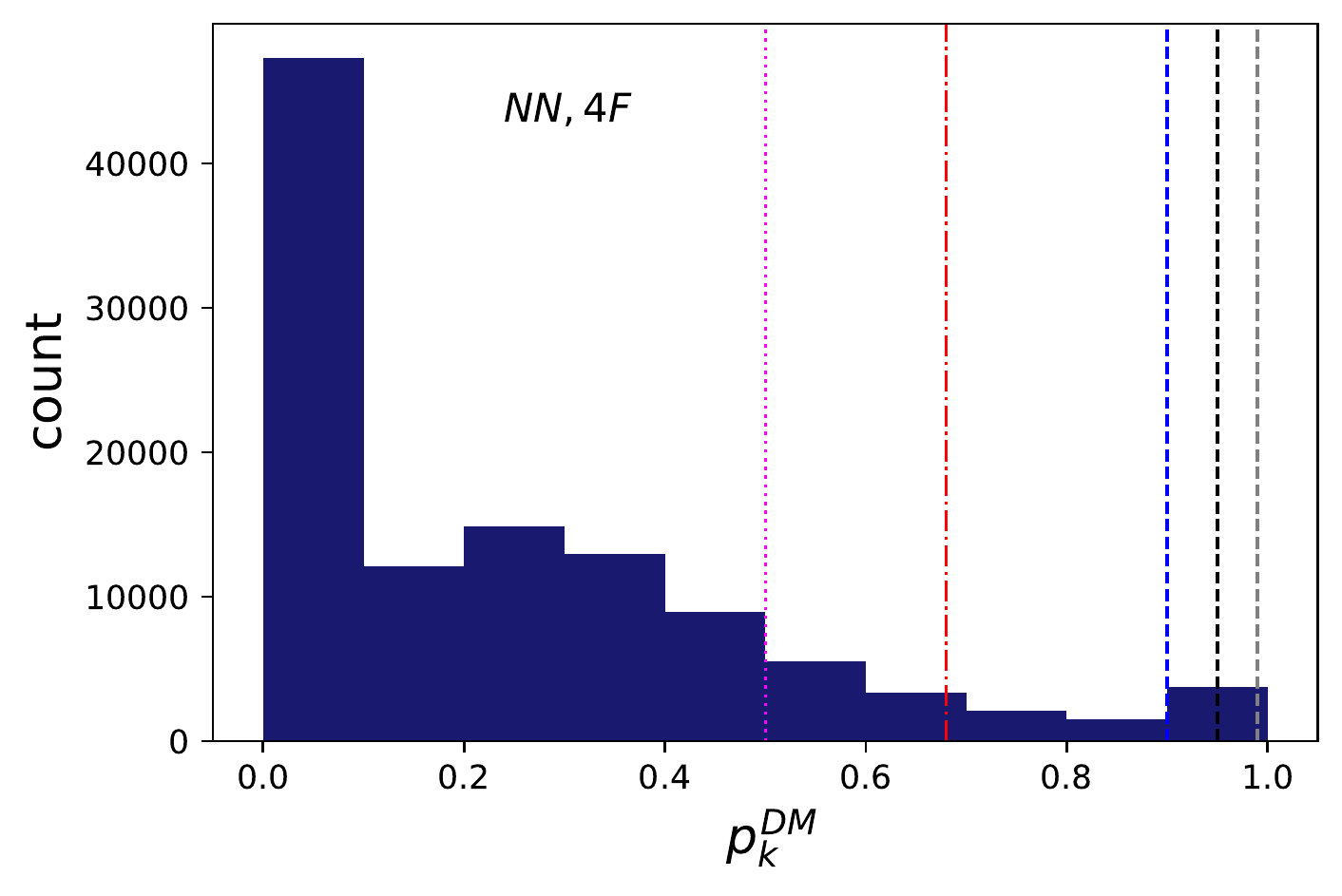}
    \caption{\protect\footnotesize{Probability distribution of the full sample of unIDs classified 100 times. The histogram as $100\times N_{\rm unids}$ entries. 
    Vertical lines correspond to different cut on $p_k^{\rm DM}$, namely 0.50 (magenta-dotted line), 0.68 (red-dotted-dashed line), 0.90 (blue-dashed line), 0.95 (black-dashed line), 0.99 (grey-dashed line). Upper panel: NN classifier in the 2F setup. Lower panel: NN classifier in the 4F setup. 
    }
    }
    \label{fig:all_unids_histo}
\end{figure}

 Among our algorithms, we select the one with the best performance - which is the NN with the 4F setup - to classify our unIDs sample and to search for prospective DM-source candidate. Nonetheless - in order to show the improvement in the classification obtained by training the NN with the inclusion of the systematic features - In Fig. \ref{fig:all_unids_histo} we show the classification results obtained from both the 2F and 4F setups. In particular, we show the distribution of probabilities $p_k^{\rm DM}$ of the full sample of unIDs to be classified as DM in each of the $k=1....100$ classification runs, corresponding to different training/testing split and/or different random seeds. Fig. \ref{fig:all_unids_histo} shows a clear trend in the astro-vs-DM classification of unIDs.
 On the one hand, many unIDs are classified with probability $30\%\leq p_k^{\rm DM}\leq 60\%$ in the 2F setup (upper panel in Fig. \ref{fig:all_unids_histo}); this peak of probabilities spreads to $0\% \leq p_k^{\rm DM}\leq 60\%$ in the 4F setup, now suggesting that those sources are most probably astrophysical sources. The improvement in such a degeneracy represents a first partial result of this work.

\begin{figure}
\begin{center}
   \includegraphics[width=\linewidth]{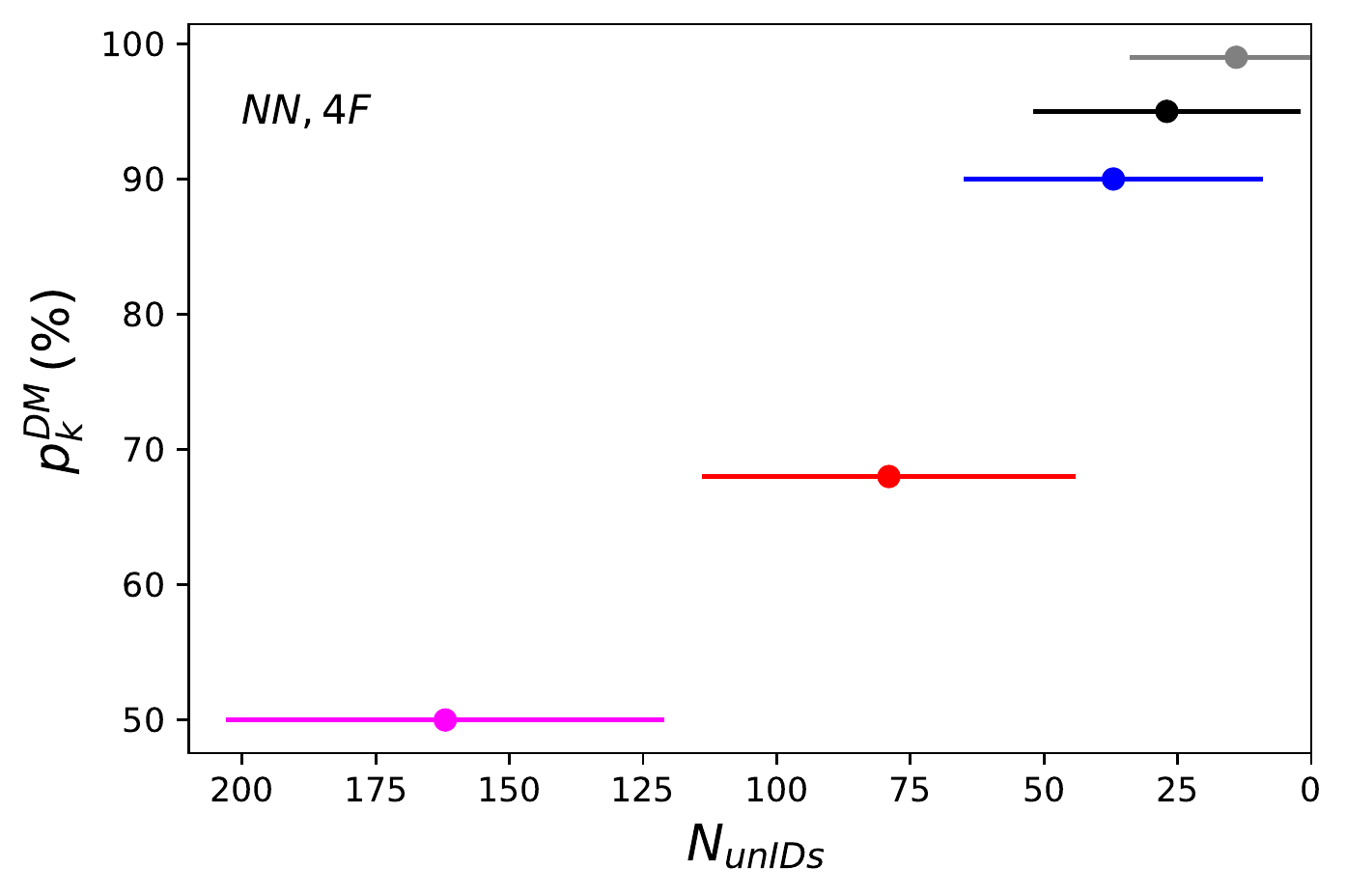}
    \caption{\protect\footnotesize{Mean number of unIDs with $p_k^{\rm DM}>0.50,0.68,0.90,0.95,0.99$ in the NN-4F classification. The error bars are calculated as the standard deviation on 100 classifications.}
    }
    \label{fig:unIDs_mean_number_4F}
    \end{center}
\end{figure}

In Tab. \ref{tab:unID_4F} and Fig. \ref{fig:unIDs_mean_number_4F}, we show the mean number of unIDs classified with $p_k^{\rm DM} \geq 50\%, 68\%, 90\%, 95\%, 99\% $ in each classification, and the standard deviation calculated on $k=1....100$ classification runs. 
\\

\begin{table}
\begin{center}
\resizebox{\columnwidth}{!}{
\begin{tabular}{|c|c|c|c|c|c|}
\hline 
\hline
Setup & $p_k^{\rm DM}\geq 50\%$ & $p_k^{\rm DM}\geq 68\%$ & $p_k^{\rm DM}\geq 90\%$ & $p_k^{\rm DM}\geq 95\%$ & $p_k^{\rm DM}\geq 99\%$\\
\hline
\hline
4F  & $162\pm 41$ & $79\pm35$ & $37\pm 28$ & $27\pm25$ & $14^{+20}_{-14}$ \\
\hline
\end{tabular} 
}
\caption{\protect\footnotesize{Result of classifying the unIDs with the NN model, for the 4F setup considered in this work. The entries represent mean and standard deviation (across the splits) of the number of unIDS (out of 1125 considered in our sample) whose prediction for the probability of being DM is greater than $50\%$, $68\%$, $90\%$, $95\%$ and $99\%$, respectively. (see also Fig. \ref{fig:unIDs_mean_number_4F}).}}
\label{tab:unID_4F}
\end{center}
\end{table}

\begin{figure}
\begin{center}
   \includegraphics[width=\linewidth]{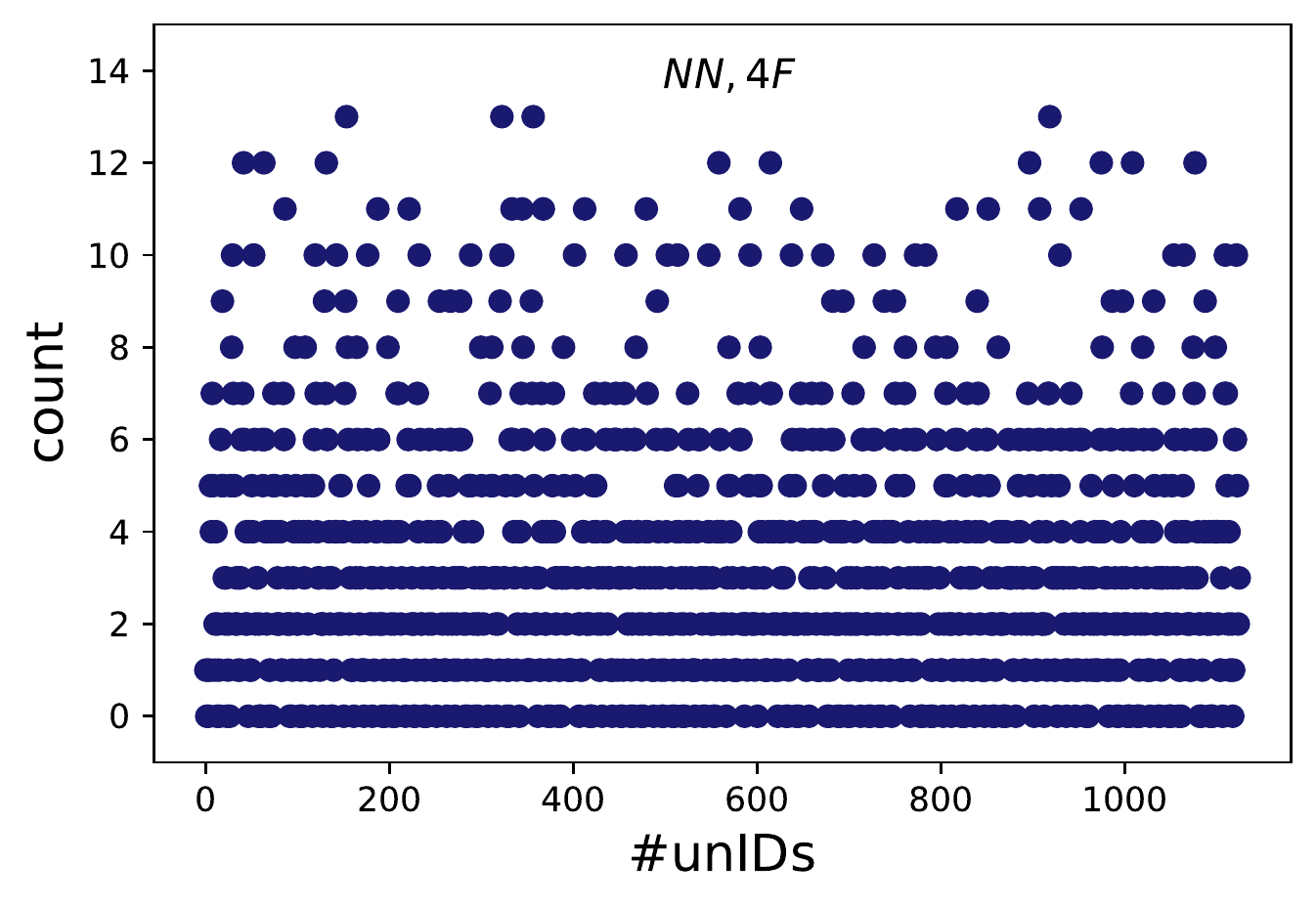}
    \caption{\protect\footnotesize{Each point reflects how many times each unIDs is classified with $p_k^{\rm DM}\geq90\%$. The best candidates in the run have been classified 13 times over 100 classifications with $p_k^{\rm DM}\geq90\%$. Nonetheless, the small counts and the statistical fluctuations with other unIDs do not allow us to claim for a robust DM candidate.}}
    \label{fig:unIDs_cand_4F}
    \end{center}
\end{figure}

Finally, although the number of unIDs classified as DM with $p_k^{\rm DM}\geq 99\%$ is compatible with zero, one may wonder which unIDs of the sample has any $p_k^{\rm DM}\geq 90\%$. 
In Fig. \ref{fig:unIDs_cand_4F} we show the counting for each unIDs to be classified with $p_k^{\rm DM} \geq 90\%$ over 100 classifications. We observe that at most 13 out of the 100 classifiers give a probability $p_k^{\rm DM}\geq 90\%$ only for a few unIDs.
Due to both such a small counting and the statistical fluctuations, it is indeed impossible to point out a specific best DM candidates among our sample of unIDs, our results being compatible with no DM sources among our unIDs sample. 

\section{Conclusions}
\label{conclusions}

The main scope of this work was to study the possibility that some of the unidentified gamma-ray point-like sources (unIDs) found at the latest Fermi-LAT catalog (4FGL) would actually shine due to WIMP DM annihilation. 

In order to do that, we first studied the differences between observed astrophysical sources (pulsars, blazars, etc) and prospective DM candidates in a 2-dimensional parameters space which have been shown to offer good discriminatory power in previous studies with astrophysical sources only. Such parameters are $E_{\rm peak}$ and $\beta$ \citep{Coronado-Blazquez:2019pny, Coronado_Blazquez2019} -defining the so-called $\beta$-plot- which among others characterise the energy spectrum of the sources when fitted with a log-parabola shape. We interpreted this problem as a standard Machine Learning (ML) binary classification task, and considered four ML models for that purpose. In doing this, we found that the above two parameters offered a limited discrimination power, while for the observed astrophysical sources there is much more information available. In particular, two additional parameters, the detection significance $\sigma_d$ and the uncertainty $\varepsilon_\beta$ associated to the $\beta$ parameter were promising for improving the classification task. In order to use those, we built fictitious values of these two additional parameters for the DM simulated data. We did it by sampling the corresponding distributions of the unIDs, under the main assumption that if WIMPs were actually present, they would show up among the unIDs population. We suggest the inclusion of these synthetic features as an heuristics to easily take into account some of the feature uncertainties in the classification algorithms. 

The ML models considered in this work are: Logistic Regression, Neural Network (NN), Naive Bayes and a particular implementation of a Gaussian Process classifier. The first three models were trained using either 2 features ($E_{\rm peak}$ and $\beta$), 3 features (including $\sigma_d$), or 4 features (including also $\varepsilon_\beta$). The 3-feature setup implemented an augmented dataset for taking into account $\varepsilon_\beta$ as part of the data itself. The 4-feature setup incorporated the information about $\varepsilon_\beta$ simply as an extra independent feature instead. The GP model on the other hand, used the same 3 features as above, while incorporating $\varepsilon_\beta$ not as an augmented dataset, but as part of the statistical model. Overall, we found that the configuration giving the best performance, in terms of classification accuracy, was the NN with 4 features (4F), giving a classification accuracy of about 93\% $\pm 0.7\%$. 

We selected such setup, NN-4F, as the one for the final task of classifying unIDs as either astrophysical sources or DM sources. We created 100 versions of such a setup, by training the model on 100 data splits (including different train+test partitions as well as random seeds for initializing the weights of the network), thus effectively having 100 predictions for every unID about the probability of being DM. We found that in most cases the predicted probabilities are smaller than 10\%, while there is a distribution extending to larger values (cf Fig. 8 lower panel). Only few unIDs are classified with a larger probability (greater than 90\%) to be DM, but only in at most 13 out of the 100 predictions (cf. Fig. 10), while subject to large statistical fluctuations. 

{\it We thus conclude that there is no significant evidence for WIMP DM among the unIDs analysed in this work.}

As a final word, we would like to remark that -although we found no DM candidates among our sample of LAT unIDs- the proposed methodology appears promising in order to include features uncertainties in classification problems, while with improving the the overall performance. In a near future we aim to apply this new methodology. In fact, the proposed methodology is completely model dependent, both on the experimental and theoretical side. On the experimental side, it depends on the characteristics of the instrument, e.g. the energy range, on the theoretical side the WIMP hypothesis could be relaxed searching for e.g. other DM candidates. 

\section*{Acknowledgments}
The work of VG and MASC was supported by the grants PGC2018-095161-B-I00, CEX2020-001007-S, PID2021-125331NB-I00 all funded by MCIN/AEI/10.13039/501100011033 and by ``ERDF A way of making Europe''.
VG's contribution to this work has been supported by \textit{Juan de la Cierva-Formaci\'on} FJCI-2016-29213 and \textit{Juan de la Cierva-Incorporaci\'on} IJC2019-040315-I grants. 
BZ has been further supported by the Programa Atracci\'on de Talento de la Comunidad de Madrid under grant no. 2017-T2/TIC-5455, from the Comunidad de Madrid/UAM “Proyecto de J\'ovenes Investigadores” grant no. SI1/PJI/2019-00294, from Spanish “Proyectos de I+D de Generacion de Conocimiento” via grant PGC2018-096646-A-I00. BZ also acknowledges the support from Generalitat Valenciana through the plan GenT program (CIDEGENT/2020/055).
The work of MASC and JCB was also supported by the Atracci\'on de Talento contracts no. 2016-T1/TIC-1542 and 2020-5A/TIC-19725 granted by the Comunidad de Madrid in Spain.
\section*{Data Availability}

The \textit{Fermi}-LAT Collaboration follows an open-access policy of data sharing: the data are provided on-line to scientific community at the following link: https://fermi.gsfc.nasa.gov/ssc/data/access/lat/10yr\_catalog/.




\bibliographystyle{mnras}
\bibliography{biblio} 




\appendix

\section{DM\texorpdfstring{$-\beta$}{} plot with systematic feature on \texorpdfstring{$\beta$.}{}}
\label{DM_beta_uncertainty}

\begin{figure}[h!]
\begin{center}
\includegraphics[width=0.45\textwidth]{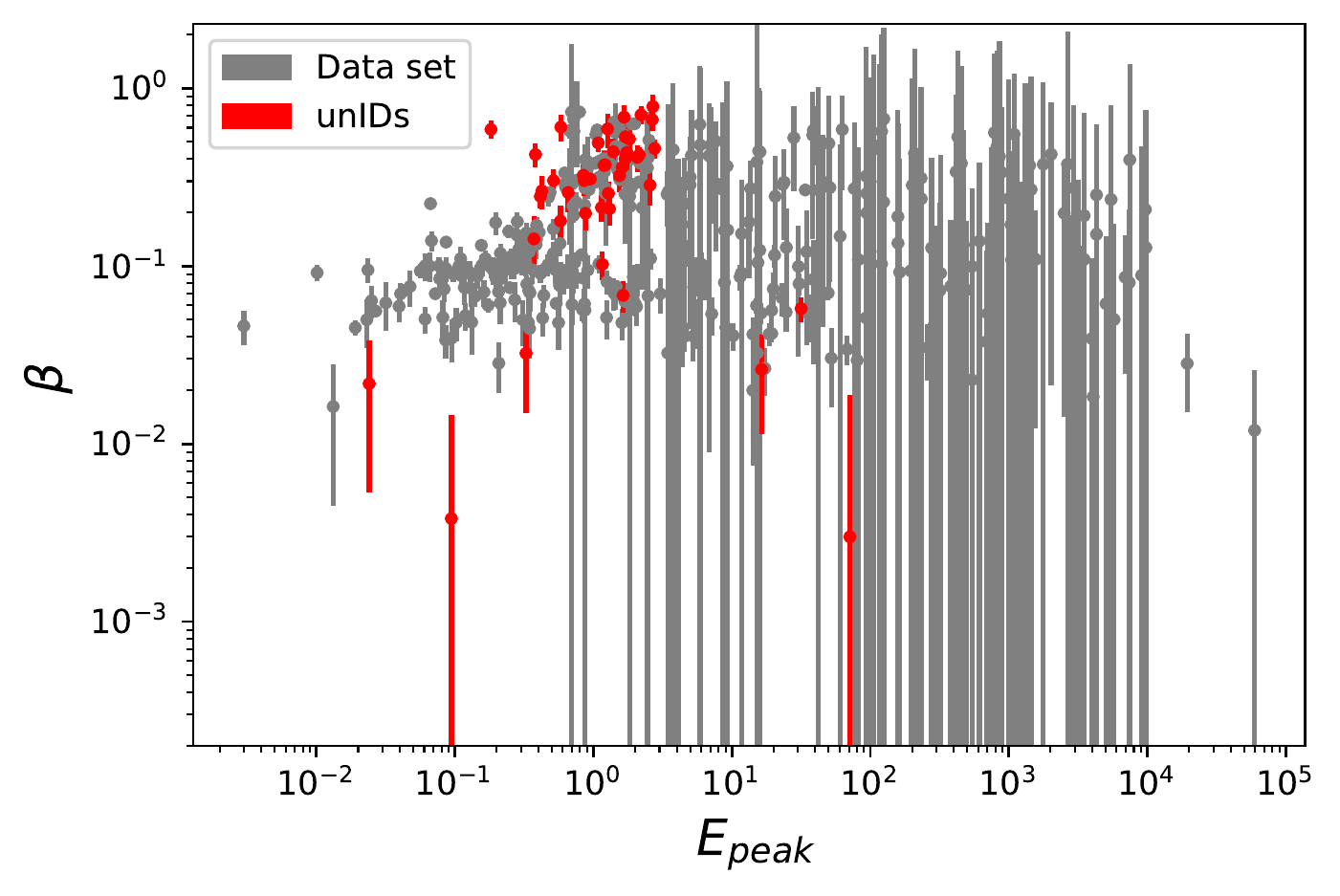}
\caption{\protect\footnotesize{Same as Fig.\ref{fig:DM_beta_plot}, by including the systematic features created for the DM data set. A cut on $\sigma \geq 20$ is applied for abetter clarity of the plot. 
}}
\label{fig:beta_plot_errors_20sigma}
\end{center}
\end{figure}

In Fig. \ref{fig:beta_plot_errors_20sigma} we show an example of the $\beta-$plot with the astrophysical and DM data set including the systematic uncertainty on $\beta$. A cut on $\sigma\geq20$ is applied for the clearness of the plot. The uncertainty on the data set is only partially correlated to the detection significance, accordingly with the observational sample of unIDs sources.

\section{Sampled Guassian distribution of \texorpdfstring{$\beta$}{} uncertainty}
\label{sec:augmented_sample}
\begin{figure}
\centering
\includegraphics[width=0.40\textwidth]{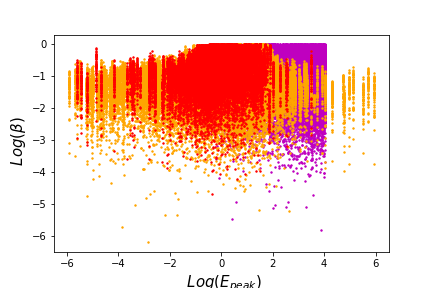}\\
\includegraphics[width=0.40\textwidth]{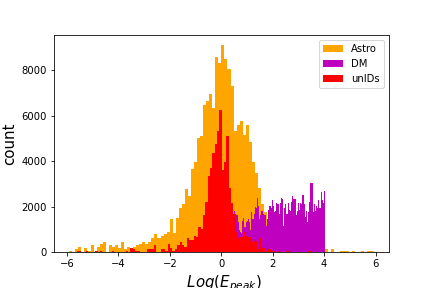}\\
\includegraphics[width=0.40\textwidth]{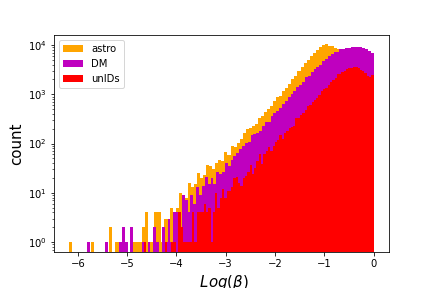}\\
\includegraphics[width=0.40\textwidth]{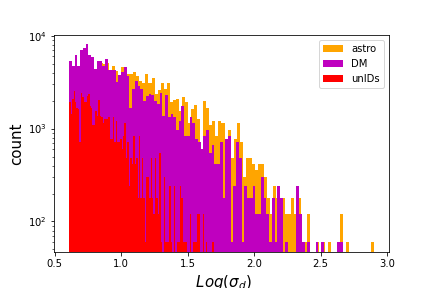}\\
\caption{\protect\footnotesize{DM-$\beta$ plot (upper panel) and histograms of the three features of the augmented data set. From the second upper to lower panel: characteristic emission energy  $E_{\rm peak}$, curvature of the spectra $\beta_{\rm sampled}$, detection significance $\sigma_{\rm d}$. In this set-up, the relative error is included as the standard deviation of a truncated Gaussian around the mean value $\beta$, i.e. we only have three features (3F-A, as explained in the \ref{sec:augmented_sample}).
}}
\label{fig:histo_sampled}
\end{figure}

In this Appendix we describe a different methodology to include the uncertainty on $\beta$ in the classification algorithm. Instead of incorporating the uncertainty of $\beta$ as an extra feature, another option is to include it implicitly in the data by creating an augmented dataset $\beta_{\rm sampled}$.The strategy here is to augment the dataset for each observation $i=0,...N$, we assume that the variable $\beta$ follows a truncated Gaussian distribution, whose mean is precisely the observed value $\beta_i$, and the standard deviation is precisely the observed value $\epsilon_{\beta_i}$, but truncated such that $0<\beta\leq 1$. We then sample $M=60$ points from such truncated Gaussian, such that we obtain an augmented dataset of size $M\cdot N$. This is an heuristic methodology for taking into account the uncertainties in the input variables, by using only three features (3F-A) $E_{\rm peak}$, $\beta_{\rm sampled}$, $\sigma_{\rm}$. We show both the "DM-$\beta$" plot and the final histograms for each feature in Fig. \ref{fig:histo_sampled}. Nonetheless, we found some issues with this methodology. Firstly, the augmented number of data makes the classification slower. Secondly, the augmented dataset imply a substantially larger overlap between the two classes of points (both in the $DM-\beta$ plot and histograms of each features, see Fig. \ref{fig:histo_sampled}), with the consequent loss in discrimination power. Thirdly, this method requires the use of only three features in the learning process. Indeed, the same algorithm applied to the unIDs classification, implies the use of only three features, with two options: 1) neglecting the $\beta_{\rm rel}$ for the unIDs and using only the mean value of $\beta$ as feature, so preventing us from using the available information contained in the unID's $\beta_{\rm rel}$; 2) including $\beta_{\rm rel}$ by augmenting the unIDs sample with the same gaussian sample methodology, which at the end will bring to an overlapping of different unIDs that would be very hard to reconstruct. For all these reasons, we do not use this methodology for the unIDs classification. 

\section{Technical details of the ML models}
\label{sec:repro}

For the sake of reproducibility, we specify in this section the implementation details of some of the ML models considered in this work.

\subsection*{Logistic Regression}

We use the implementation of LR as given in the python library  {\tt Scikit-learn} \citep{scikit-learn}, indeed class: \\
\\
\textit{sklearn.linear$\_$model.LogisticRegression (penalty='l2', *, dual=False, tol=0.0001, \, 
C=1.0, fit$\_$ intercept=True, intercept$\_$ scaling=1, class\_ weight=None, random$\_$ state=None, solver='lbfgs', max$\_$ iter=100, multi$\_$ class='auto', verbose=0, warm$\_$ start=False, n$\_$ jobs=None, l1$\_$ ratio=None)}


\subsection*{Neural Network}

We use the implementation of NN as given in the python library  {\tt Scikit-learn} \citep{scikit-learn}. Specifically, we use the {\tt MLPClassifier} model, with the following setup:\\

\textit{sklearn.neural$\_$network.MLPClassifier(hidden$\_$layer$\_$sizes=(21,), activation='relu', *, solver='adam', alpha=0.0,batch$\_$size=120, learning$\_$rate='constant',learning$\_$rate$\_$init=0.0015,power$\_$t=0.5, max$\_$iter=1000,shuffle=True,random$\_$state=None,tol=0.0001, verbose=False,warm$\_$start=False,momentum=0.9, nesterovs$\_$momentum=True,early$\_$stopping=False, validation$\_$fraction=0.1,beta$\_$1=0.9, beta$\_$2=0.999,epsilon=1e-08, n$\_$iter$\_$no$\_$change=10,max$\_$fun=15000)}.\\
\\

\begin{figure}
    \centering
    \includegraphics[width=\linewidth]{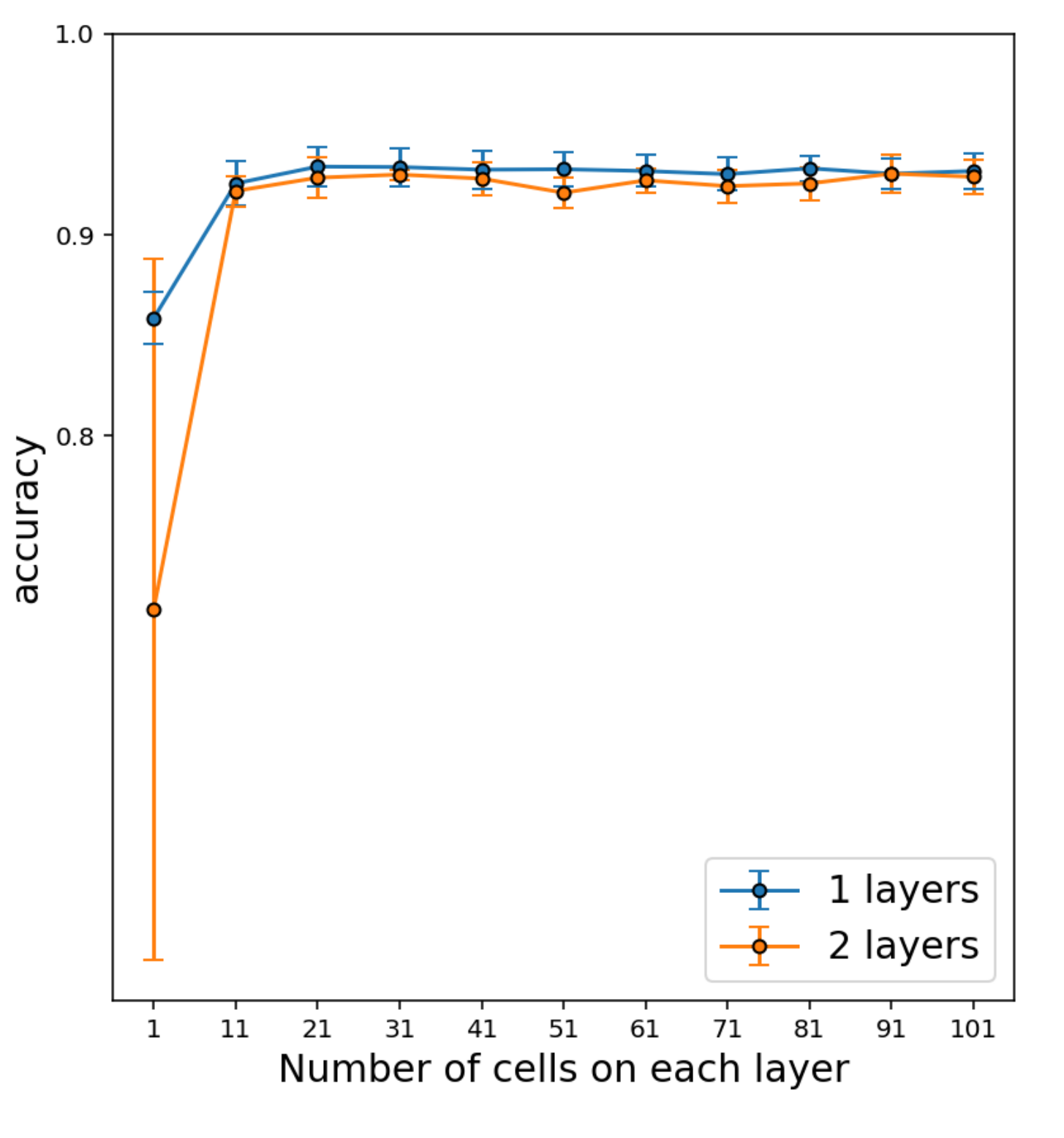}
    \includegraphics[width=\linewidth]{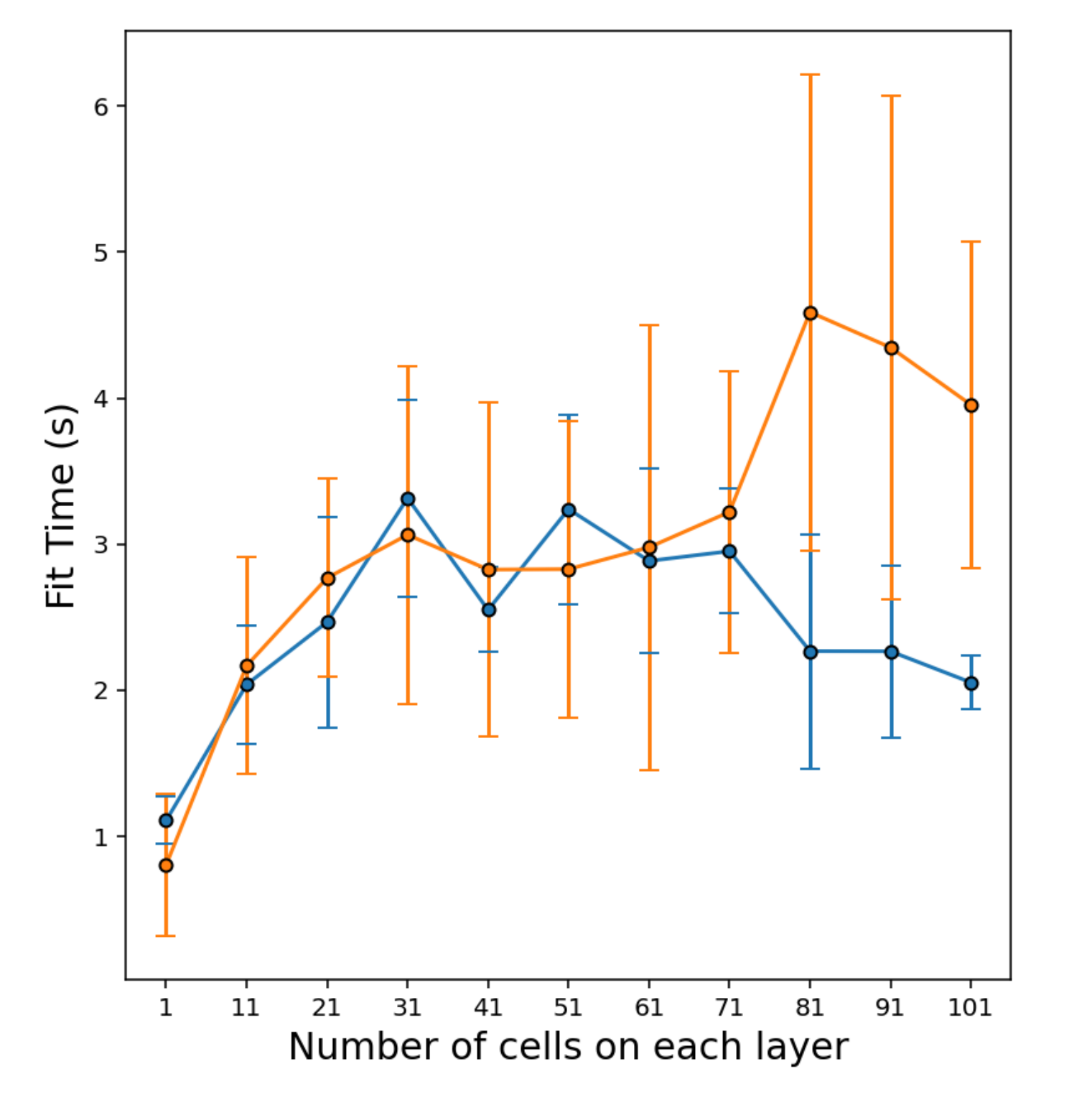}
    \caption{\protect\footnotesize{Performance comparing 1-hidden-layer and  2-hidden-layer neural network for the 4F data set. Upper panel: Overall Accuracy as a function of the number of neurons in each layer. Lower panel: Time of the learning procedure as a function the number of neurons in each layer. In the following, we use 1 layer and 21 neurons.}}
    \label{fig:layers}
\end{figure}

In Fig. \ref{fig:layers} we show the performance of the NN for one and two layers with different number of neurons. We choose to use a 1-layer configuration with 21 neurons, due to the combination of good accuracy (upper panel) and low fitting time (lower panel).

\subsection*{Gaussian Process}
\label{NPMGP}

As for the previous models, in this case we also work in log-space of the attributes. However, care must be taken in this case, since then the uncertainties of $\log\beta$ should be computed properly. This is a standard procedure, which however is depicted next:

The idea is to compute the uncertainties of a new variable $\ell\equiv \log\beta$, where $\beta \sim {\cal N}(\beta|\mu_\beta, \sigma_\beta)$, while the given uncertainties in $\beta$ are assumed to be precisely its standard deviation, $\sigma_\beta$. In order to do this, we take as the $\ell$'s uncertainties to be the square root of its variance, computed from its own pdf: 

\begin{equation}
p(\ell|\mu_\beta,\sigma_\beta) = e^\ell 
{\cal N}(e^\ell|\mu_\beta,\sigma_\beta)~.
\end{equation}

We have adapted to our case one of the variants of the Gaussian Process (GP) models presented in \citep{2020arXiv200110523V}: the NIMGP$_{\rm NN}$ model, which is roughly described next, while for further details we refer the reader to \citep{2020arXiv200110523V}.

The NIMGP$_{\rm NN}$ model assumes a likelihood for the label $y_i$ of the $i$-th point as follows: 

\begin{eqnarray}
p(y_i|\mathbf{f}_i) &=& (1-\epsilon) \prod_{c\neq y_i} \Theta\left(f^{y_i}(\mathbf{x}_i) - f^c(\mathbf{x}_i)\right) + \\ \nonumber
&&\frac{\epsilon}{C-1} \left[1 - \prod_{c\neq y_i} \Theta\left(f^{y_i}(\mathbf{x}_i) - f^c(\mathbf{x}_i)\right)\right]\,,
\label{eq:likelihood_function}     
\end{eqnarray}
where ${\bf f}_i \equiv\{f^c({\bf x}_i)\}$, $c=1,..,C$, are the corresponding values of the GP for all $C$ classes, evaluated at the latent inputs ${\bf x}_i$ (see below). We account for the possibility of having mislabelled classes by having a small probability $\epsilon=0.001$ for mislabelling. Finally, $\Theta(\cdot)$ is the Heaviside step function. Note that this is not the typical likelihood considered in popular classification tasks, which correspond to the cross-entropy loss function. However, this is a common choice in the GP classification context, with the added value of accounting for mislabelling errors, something which is typically not taking into account in the cross-entropy setup.   

In NIMGP$_{\rm NN}$ it is assumed that the observed input ${\bf\tilde x}_i$ is a noisy realisation of the true (but latent) input ${\bf x}_i$, according to the distribution:
\begin{equation}
    p({\bf\tilde x}_i|{\bf x}_i)
    ={\cal N}({\bf\tilde x}_i|{\bf x}_i, \sigma_i)~.
\end{equation}

As it is well known, the posterior distribution $p({\bf f}|{\bf y})$ of the GP values ${\bf f}$ has a computational cost of ${\cal O}(N^3)$, where $N$ is the number of points, so this setup is not scalable to very large datasets. For that reason the NIMGP$_{\rm NN}$ adopts a ``Sparse GP'' configuration where inference is done only in a subset ${\bf u}$ of GP values at some given inputs, called in the literature ``inducing points'', and there are $M<N$ of them. Consequently, the latent variables of the model can be grouped in three matrices: i) ${\bf F}$, the $N\times C$ matrix of process values ${\bf f}_i$ at the datapoints, ii) ${\bf U}$, the $M\times C$ matrix of process values ${\bf u}_j$ at the inducing points, and iii) ${\bf X}$, the $N\times D$ matrix of latent inputs ${\bf x}_i$. 

The posterior distribution for the above latent variables is intractable, as typical in Bayesian inference, and NIMGP$_{\rm NN}$ approximates it by Variational Inference, where the approximate distribution $q({\bf x}_i)$ is taken as: 

$$q(\mathbf{x}_i)=\mathcal{N}\big(\mathbf{x}_i|\boldsymbol{ \mu}_\theta(\mathbf{\tilde x}_i,y_i),\mathbf{V}_{\theta}(\mathbf{\tilde x}_i,y_i)\big) $$

where both $\boldsymbol{\mu}_\theta(\tilde{\mathbf{x}}_i,y_i)$ and $\mathbf{V}_\theta(\tilde{\mathbf{x}}_i,y_i)$ are obtained as the output of a neural network
with parameters $\theta$. 

The final scope in Bayesian inference is to compute the predictive distribution $p(y_*|{\bf x}_*, {\cal D})$,
for the class $y_*$ at a new input ${\bf x}_*$, given the already observed (training) data ${\cal D}$, which in the case of GP binary classifier is given by:
\begin{equation}
p(y_*|{\bf x}_*, {\cal D}) = \int df_* p(y_*|f_*) p(f_*|{\bf x}_*,{\cal D})~,   
\end{equation}
where $f_*$ is the process value at the input ${\bf x}_*$, while
\begin{equation}
p(f_*|{\bf x}_*,{\cal D}) = 
\int d{\bf f}~ p(f_*|{\bf x}_*,{\bf f}) p({\bf f}|{\cal D})
~,
\end{equation}
being $p({\bf f}|{\cal D})$ the posterior distribution of the process values at all the training points.
\newline\newline\noindent
As a final note, the NIMGP$_{\rm NN}$ model is written in {\tt python 2} with {\tt tensorflow 1} (see github repository at \citet{2020arXiv200110523V}). 

\section{Number of folds}
\begin{figure}
    \centering
    \includegraphics[width=1\linewidth]{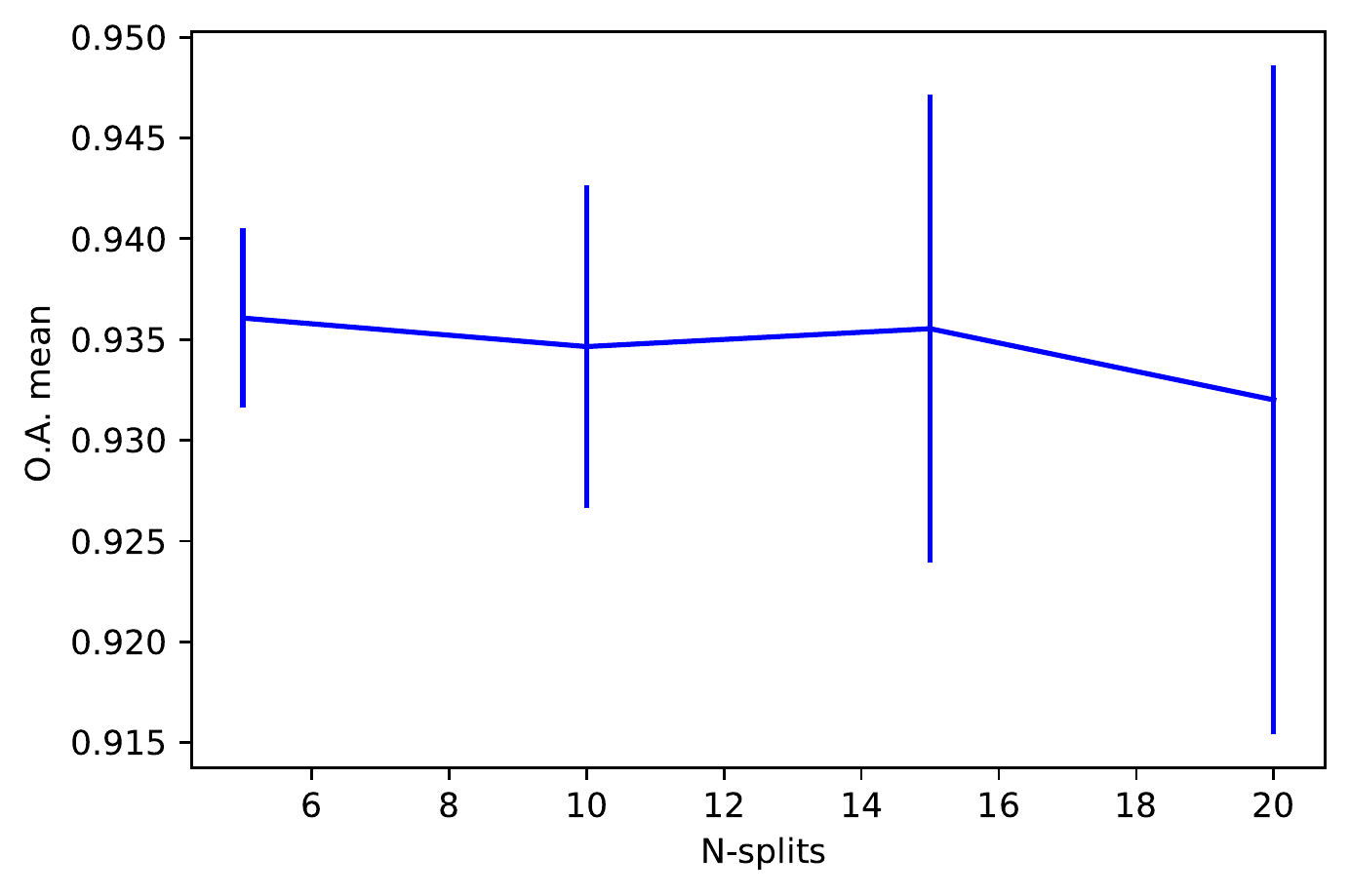}
\caption{\protect\footnotesize{Performance comparing the OA of the NN for different N-folds, for the 4F data set. See Sec. \ref{sec:class_acc} for details.}
    }
    \label{fig:OA_std_splits}
\end{figure}

For the LR and NN, we use the Repeated Stratified K-Fold cross validator, class \texttt{RepeatedStratifiedKFold(n$\_$splits=5, n$\_$repeats=20}) defined in \texttt{Scikit-learn} \citep{scikit-learn}. By splitting the data in 5-folds, we take $80\%$ of data for the training set and $20\%$ of data for the testing set. This choice allows us to preserve the independence of the 5 testing sets, i.e. without any repetition of same data. In order to preserve this characteristic, the ratio of testing/training data decreases by increasing the number of folds, while the accuracy of the classification decreases (Fig. \ref{fig:OA_std_splits}). Such a split, is repeated 20 times with different random seeds. In this way we have a total of 100 classifications, which allow us to get the results with a reliable statistical uncertainty.
\label{sec:Nfolds}

\begin{figure}
    \centering
    \includegraphics[width=1\linewidth]{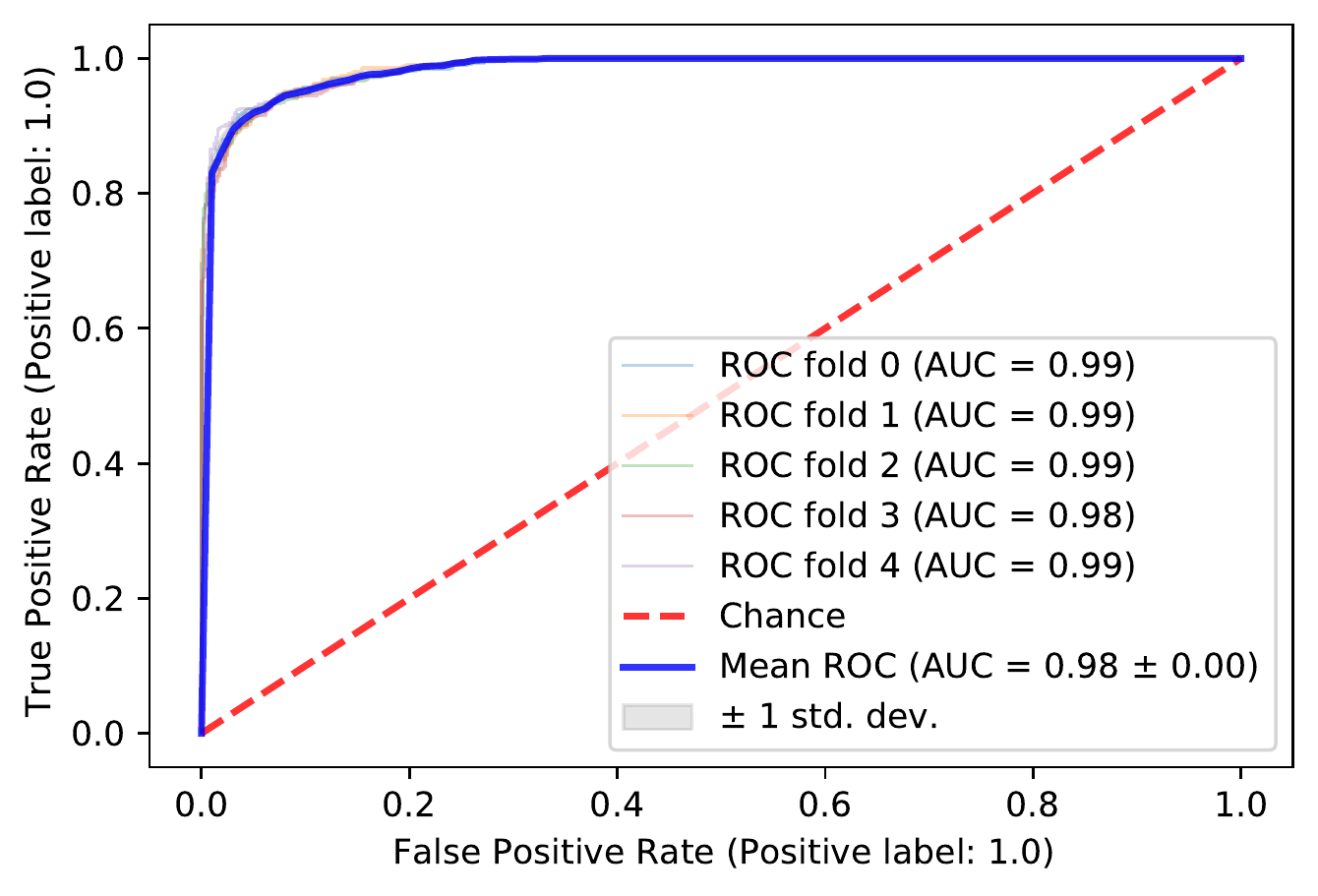}
\caption{\protect\footnotesize{Performance comparing the ROC (Receiver Operating Characteristics) and AUC (Area Under The Curve) of the NN for different 5-folds, for the 4F data set. A model with perfect performance will have AUC = 1, which means it has a good measure of separability. The red dashed line indicates the worst ROC situation, where the AUC = 0.5, the model has no discrimination capacity to distinguish between positive class and negative class.}
    }
    \label{fig:ROC_AUC_4F}
\end{figure}

The uncertainty related with the number of folds is also shown in Fig. \ref{fig:ROC_AUC_4F}, where we show the ROC (Receiver Operating Characteristics) and AUC (Area Under The Curve) of the NN for different 5-folds, for the 4F data set.

\section{Classification with 2-features}
\label{sec:2F_class}

\begin{figure}
    \centering
    \includegraphics[width=\linewidth]{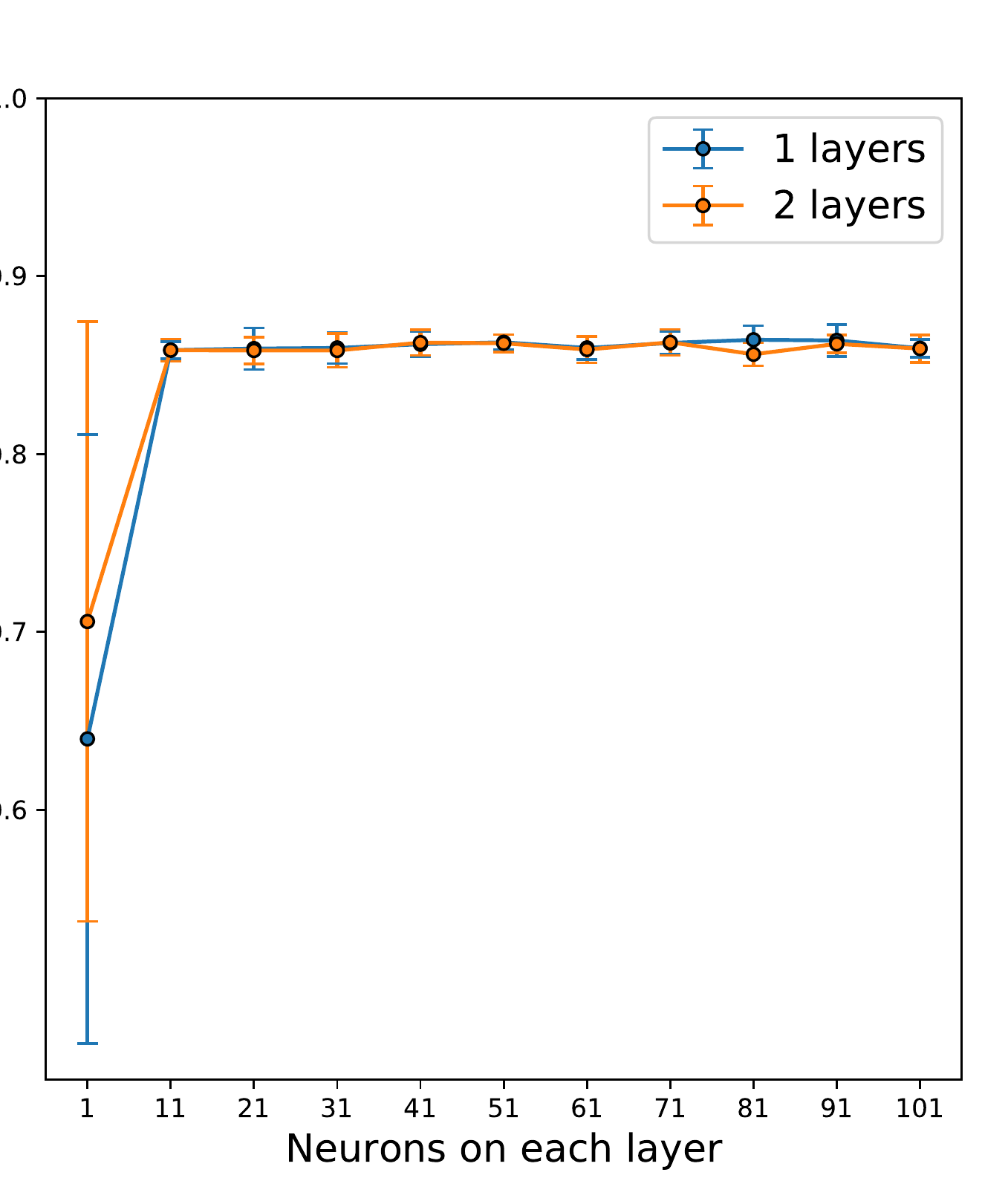} \includegraphics[width=\linewidth]{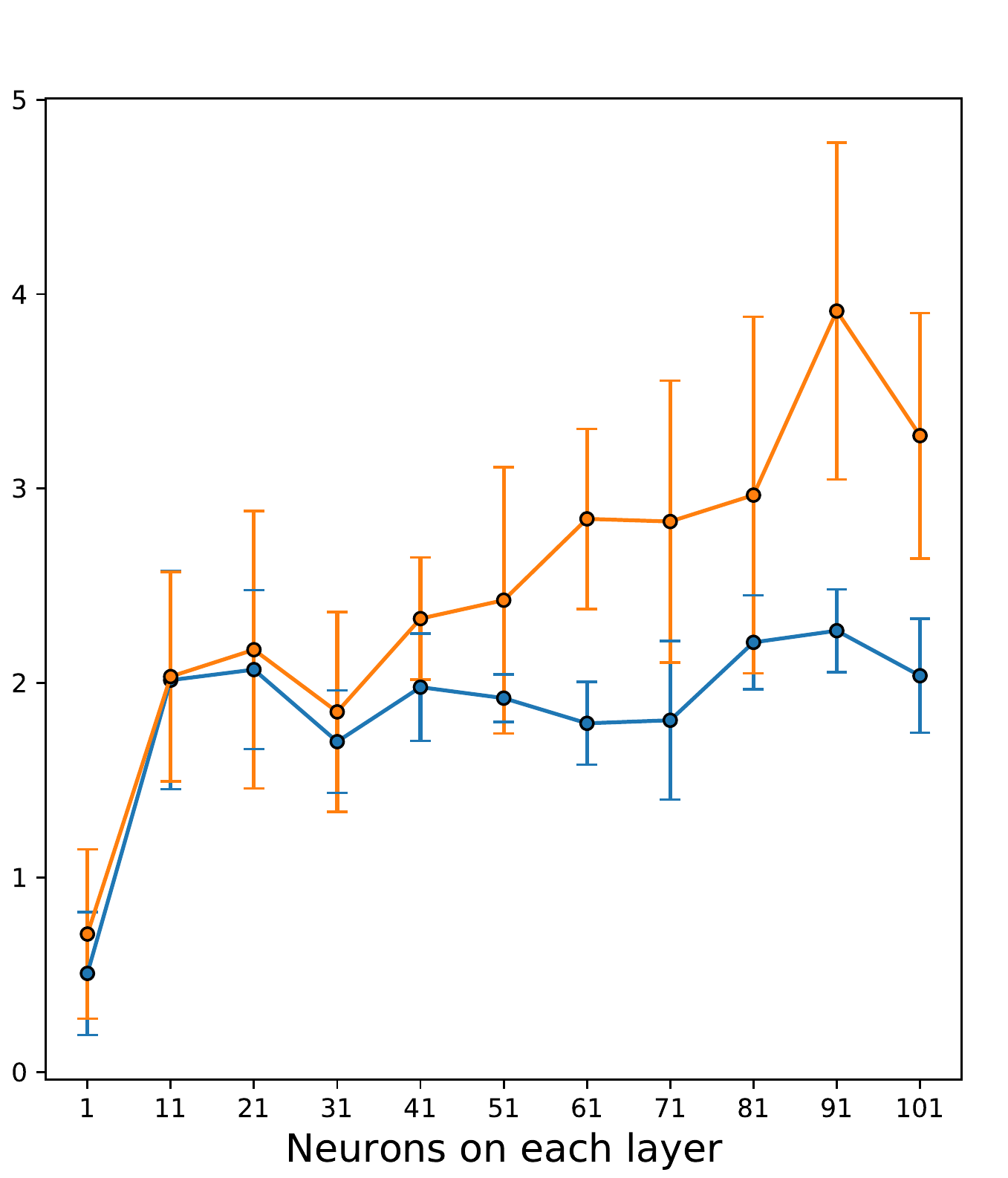}
    \caption{\protect\footnotesize{Same as Fig. \ref{fig:layers} for the 2F classification.}}
    \label{fig:layers_2F}
\end{figure}

\begin{figure}
    \centering
    \includegraphics[width=1\linewidth]{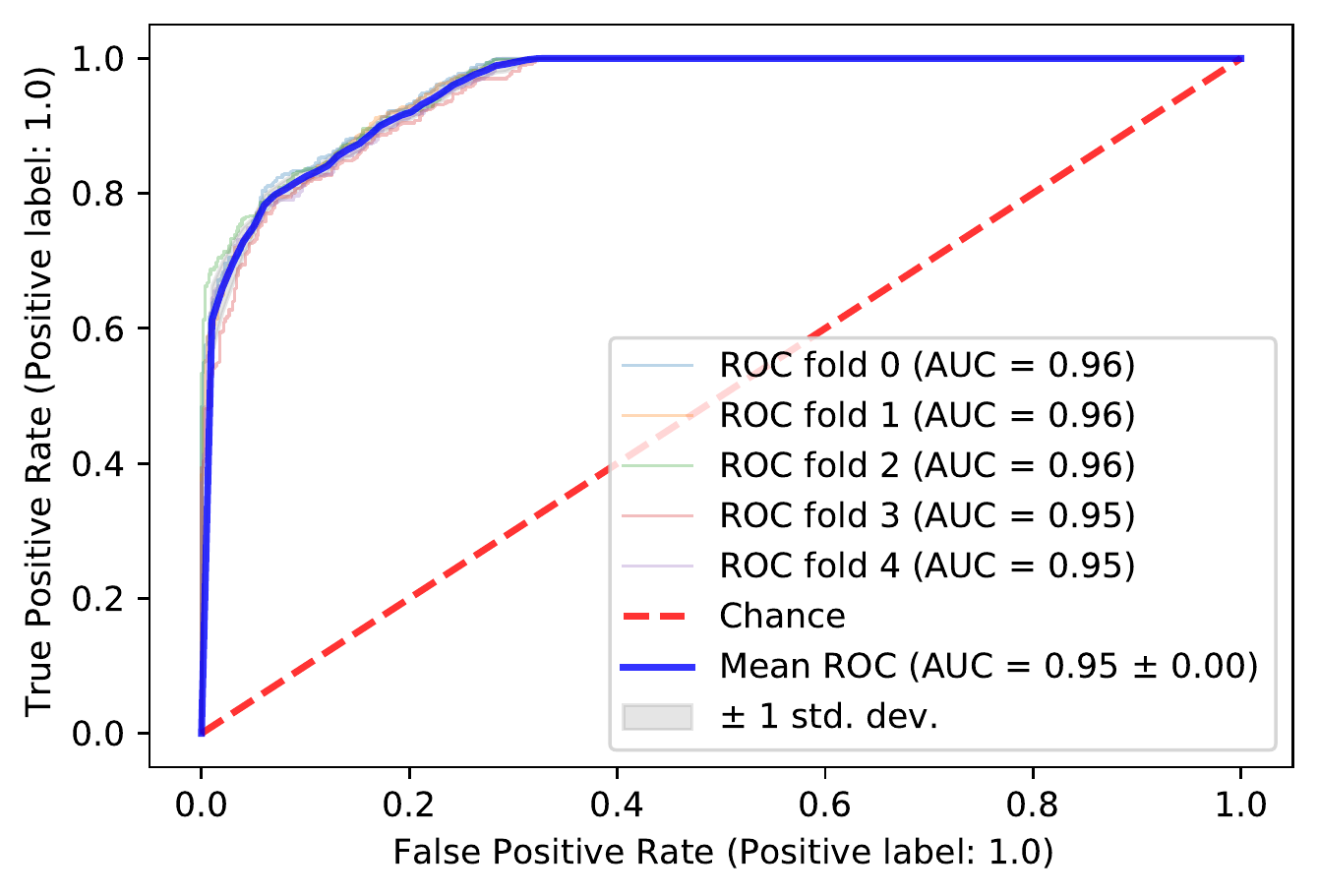}
\caption{\protect\footnotesize{Same as in Fig. \ref{fig:ROC_AUC_4F} for the classification without systematic features.}
    }
    \label{fig:ROC_AUC_2F}
\end{figure}

In Fig.s \ref{fig:layers_2F} and \ref{fig:ROC_AUC_2F} we show the same analysis of the algorithm performances for the classification with 2-features, indeed without taking into account the systematic features. The performance of the algorithm improves from 2F to 4F both in terms of overall accuracy and ROC/AUC. 

\begin{figure}
\begin{center}
\includegraphics[width=\linewidth]{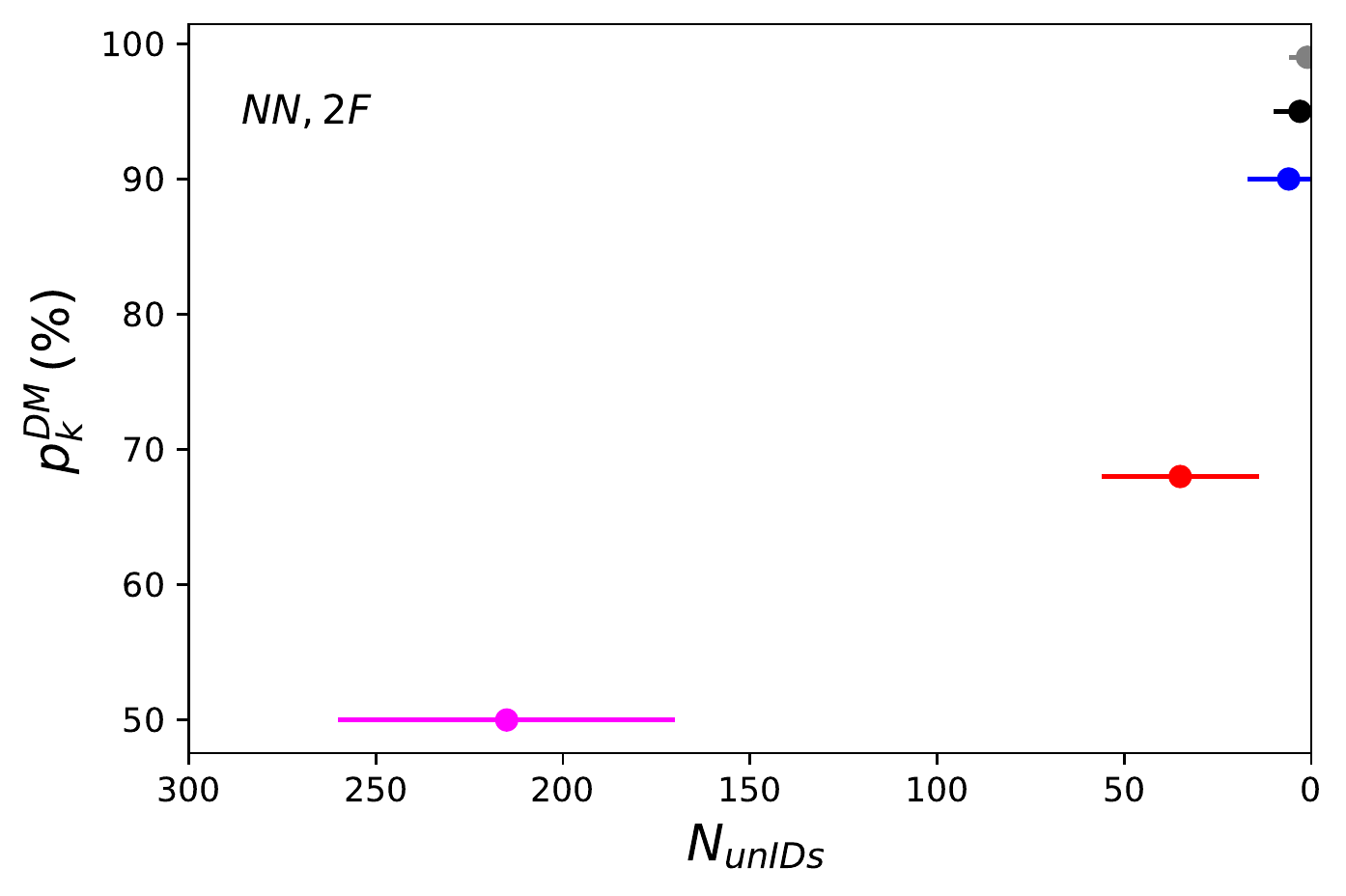}
\includegraphics[width=\linewidth]{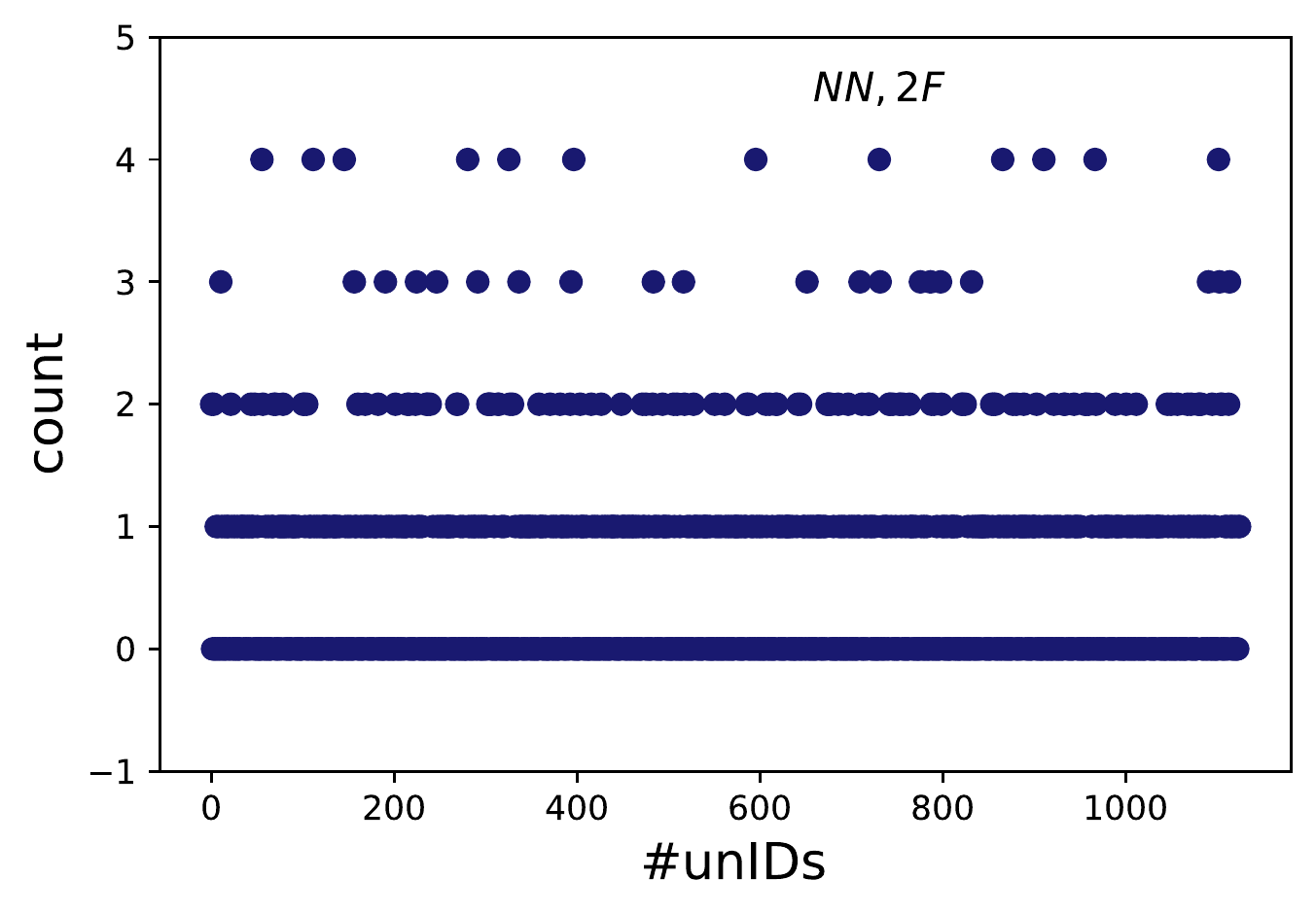}
\caption{\protect\footnotesize{Same as Fig.s \ref{fig:unIDs_mean_number_4F}, \ref{fig:unIDs_cand_4F} for the NN-2F classification. 
}
}
\label{fig:unids_class_2F}
\end{center}
\end{figure}

In Tab \ref{tab:unID_2F} and Fig. \ref{fig:all_unids_histo} of this Appendix we show the results of the unIDs classification for the NN in the 2F setup. In Fig \ref{fig:histo} in the main text, we show the probability distribution for the full set of unIDs and 100 classification runs. The improvement in the overall classification behaviour of the 4F setup is visible by comparing this figure with the same figure for the 4F setup, presented in the main text. In the upper panel in Fig. \ref{fig:unids_class_2F}, we show the mean number of unIDs classified with $p_k^{\rm DM}>0.5, 0.68, 0.90, 0.95, 0.99$ and their standard deviation. Finally, in the last panel we showed the count for each unIDs to be classified with $p_k^{\rm DM} \geq 0.90$. The best candidates are classified 5 times over 100 for the 2F setup (and 13 times over 100 in the 4F setup presented in the main text). Although the number of unIDs with $p_k^{DM}\geq  50\%$ decreases a $27\%$ from the 2F to the 4F setups, the number of unIDs with $p_k^{DM}\geq  68\%$ increases by $50\%$. Yet the number of unIDs with $p_k^{DM}\geq  99\%$ is compatible with zero for both the 2F and 4F setups. As in the 4F setup, the statistical fluctuations prevent us from claiming any robust DM candidate among the unIDs of the 4FGL Fermi-LAT catalogue.  

\begin{table}
\begin{center}
\resizebox{\columnwidth}{!}{
\begin{tabular}{|c|c|c|c|c|c|}
\hline 
\hline
Setup & $p_k^{\rm DM}\geq 50\%$ & $p_k^{\rm DM}\geq 68\%$ & $p_k^{\rm DM}\geq 90\%$ & $p_k^{\rm DM}\geq 95\%$ & $p_k^{\rm DM}\geq 99\%$\\
\hline
\hline
2F  & $215\pm 45$ & $35\pm 21$ & $6^{+10}_{-6}$ & $3^{+7}_{-3}$ &$1^{+5}_{-1}$\\
\hline
\hline
\end{tabular}
}
\caption{\protect\footnotesize{Same as Tab. \protect\ref{tab:unID_4F} for the NN in the 2F setup. See also the upper panel in Fig. \protect\ref{fig:unids_class_2F}}.}
\label{tab:unID_2F}
\end{center}
\end{table}

\bsp	
\label{lastpage}
\end{document}